\documentclass[11pt,a4paper]{article}
\pdfoutput=1
\usepackage{amsfonts}
\usepackage{amssymb}
\usepackage[centertags]{amsmath}
\usepackage{graphicx}
\usepackage{hyperref}
\usepackage{booktabs}
\usepackage{theorem}
\usepackage{array}
\usepackage{epsfig}
\usepackage{colordvi}
\usepackage{color}
\usepackage{ulem}
\usepackage[small,bf]{caption}
\usepackage{cite}
\usepackage{color}

\def\1{\mathbf{1}}
\def\3{\mathbf{3}}
\def\2{\mathbf{2}}

\newcommand{\meff}{\mbox{$\left|  \langle \!\,  m  \,\!  \rangle \right| $}}
\newcommand{\betabeta}{\mbox{$(\beta \beta)_{0 \nu}  $}}
\def\ltap{\ \raisebox{-.4ex}{\rlap{$\sim$}} \raisebox{.4ex}{$<$}\ }
\def\gtap{\ \raisebox{-.4ex}{\rlap{$\sim$}} \raisebox{.4ex}{$>$}\ }

\allowdisplaybreaks[1]

\usepackage{a4wide}

\newcommand{\bec}{\begin{cases}}
\newcommand{\eec}{\end{cases}}
\newcommand{\beq}{\begin{equation*}}
\newcommand{\eeq}{\end{equation*}}
\newcommand{\be}{\begin{equation}}
\newcommand{\ee}{\end{equation}}
\newcommand{\ba}{\begin{eqnarray}}
\newcommand{\ea}{\end{eqnarray}}

\DeclareMathOperator{\ci}{\text{i}}

\makeatletter

\newcommand{\Rmnum}[1]{\expandafter\@slowromancap\romannumeral #1@}
\makeatother

\begin{document}

\begin{titlepage}

\vspace*{-15mm}
\begin{flushright}
SISSA 20/2014/FISI\\
IPMU14-0104\\
arXiv:1405.6006
\end{flushright}
\vspace*{0.7cm}

\begin{center}
{\bf\LARGE {Predicting the Values of the Leptonic CP Violation 
Phases in Theories with Discrete Flavour Symmetries}}\\
[8mm]
S. T. Petcov$^{\, a,b,}$ \footnote{Also at:
 Institute of Nuclear Research and Nuclear Energy,
  Bulgarian Academy of Sciences, 1784 Sofia, Bulgaria.} 
\\[1mm]
\end{center}
\vspace*{0.50cm}
\centerline{$^{a}$ \it SISSA/INFN, Via Bonomea 265, I-34136 Trieste, Italy }
\vspace*{0.2cm}
\centerline{$^{b}$ \it Kavli IPMU (WPI), University of Tokyo,\\
5-1-5 Kashiwanoha, 277-8583 Kashiwa, Japan}
\vspace*{1.20cm}

\begin{abstract}
\noindent
Using the fact that the neutrino mixing matrix 
$U = U^\dagger_{e}U_{\nu}$, where $U_{e}$ and $U_{\nu}$ 
result from the diagonalisation of the charged lepton 
and neutrino mass matrices, we consider a 
number of 
forms of  $U_{\nu}$ associated with a variety 
of discrete symmetries:
i) bimaximal (BM) and ii) tri-bimaximal (TBM) forms, 
the forms corresponding iii) to the conservation of the 
lepton charge $L' = L_e - L_\mu - L_{\tau}$ (LC), 
iv) to golden ratio type A (GRA) mixing,  
v) golden ratio type B (GRB) mixing, 
and vi) to hexagonal (HG) mixing.  
Employing the minimal form of $U_e$, in terms of angles and 
phases it contains, that can provide the requisite 
corrections to $U_{\nu}$ so that 
reactor, atmospheric and solar neutrino mixing angles 
$\theta_{13}$, $\theta_{23}$ and  $\theta_{12}$ 
have values compatible with the current data,
including a possible sizable deviation of $\theta_{23}$ 
from $\pi/4$, we discuss the possibility to obtain
predictions for the CP violation phases 
in the 
neutrino mixing matrix. 
Considering the ``standard ordering'' 
of the the 12 and the 23 rotations 
in $U_e$ and following the approach 
developed in \cite{Marzocca:2013cr} 
we derive predictions for the Dirac phase 
$\delta$ and the rephasing invariant $J_{\rm CP}$
in the cases of GRA, GRB and HG forms of $U_{\nu}$
(results  for the TBM and BM (LC) forms 
were obtained in \cite{Marzocca:2013cr}). 
We show also that under rather general conditions  
within the scheme considered the values of 
the Majorana phases in the PMNS matrix 
can be predicted for each of the 
forms of $U_{\nu}$ discussed. We give examples 
of these predictions and of their implications 
for neutrinoless double beta decay.
In the GRA, GRB and HG cases, 
as in the TBM one, relatively large CP 
violation effects in neutrino oscillations 
are predicted ($|J_{CP}| \sim (0.031 - 0.034)$).
Distinguishing between the TBM, 
BM (LC), GRA, GRB and HG forms of $U_{\nu}$  requires 
a measurement of $\cos\delta$ or a 
relatively high precision measurement of $J_{\rm CP}$. 

\end{abstract}

\vspace{0.5cm}

\end{titlepage}
\setcounter{footnote}{0}

\newpage

\section{Introduction}

Determining the status of the CP symmetry in the lepton sector 
is one of the highest priority principal goals of the program of 
future research in neutrino physics 
(see, e.g., \cite{PDG2014,LBLFuture13}).
As in the case of the quark sector, the CP symmetry 
can be violated in the lepton sector by the presence of 
physical phases in the Pontecorvo, Maki, Nakagawa and Sakata (PMNS) 
neutrino mixing matrix.
In the case of 3-neutrino mixing and massive 
Majorana neutrinos we are going to consider 
\footnote{All compelling data on 
neutrino masses, mixing and oscillations are 
compatible with the existence of mixing of three 
light massive neutrinos $\nu_i$, $i=1,2,3$, 
in the weak charged lepton current
(see, e.g., \cite{PDG2014}).
It follows also from the data that the masses 
$m_i$ of the three light neutrinos $\nu_i$ do 
not exceed approximately
1 eV, $m_i \ltap 1$ eV, i.e., they are 
significantly smaller than the 
masses of the charged leptons and quarks.},
the $3\times 3$ unitary  PMNS 
matrix $U_{\rm PMNS}\equiv U$ contains, as is well known,  
one Dirac and two Majorana \cite{BHP80} 
CP violation (CPV) phases which can be the 
source of CP violation in the lepton sector. 
In the widely used standard parametrisation \cite{PDG2014} 
of the PMNS matrix we also are going to employ,  
$U_{\rm PMNS}$ is expressed in terms of the solar, 
atmospheric and reactor neutrino mixing 
angles $\theta_{12}$,  $\theta_{23}$ and
$\theta_{13}$, respectively, and the Dirac and
Majorana CPV phases, as follows:
\begin{equation}
U= VQ\,,~~~
Q = {\rm diag}(1, e^{i \frac{\alpha_{21}}{2}}, e^{i \frac{\alpha_{31}}{2}})\,, 
\label{eq:VQ}
\end{equation}
%
where $\alpha_{21,31}$ 
are the two Majorana CPV 
phases and $V$ is a CKM-like matrix, 
\begin{equation} 
\begin{array}{c}
\label{eq:Vpara}
V = \left(\begin{array}{ccc} 
 c_{12} c_{13} & s_{12} c_{13} & s_{13} e^{-i \delta}  \\[0.2cm] 
 -s_{12} c_{23} - c_{12} s_{23} s_{13} e^{i \delta} 
 & c_{12} c_{23} - s_{12} s_{23} s_{13} e^{i \delta} 
 & s_{23} c_{13} 
\\[0.2cm] 
 s_{12} s_{23} - c_{12} c_{23} s_{13} e^{i \delta} & 
 - c_{12} s_{23} - s_{12} c_{23} s_{13} e^{i \delta} 
 & c_{23} c_{13} 
\\ 
  \end{array}    
\right)\,. 
\end{array} 
\end{equation}
%
\noindent 
In eq.~(\ref{eq:Vpara}), 
$\delta$ is the Dirac CPV phase,  
$0 \leq \delta \leq 2\pi$,
we have used the standard notation
$c_{ij} = \cos\theta_{ij}$, 
$s_{ij} = \sin\theta_{ij}$, and   
$0 \leq  \theta_{ij} \leq \pi/2$.
In what concerns the Majorana CPV phases,
for the purpose of the present study it is sufficient 
to consider that they vary in the intervals 
$0 \leq \alpha_{21,31} \leq 2\pi$~
\footnote{
One should keep in mind, however, that 
in the case of the seesaw mechanism of neutrino 
mass generation the Majorana phases 
$\alpha_{21}$ and $\alpha_{31}$ 
vary in the interval \cite{EMSPEJP09}
$0 \leq \alpha_{21,31} \leq 4\pi$. 
The interval beyond $2\pi$, 
$2\pi \leq \alpha_{21,31} \leq 4\pi$, is relevant, 
e.g., in the calculations of the baryon asymmetry within 
the leptogenesis scenario \cite{EMSPEJP09},  
in the calculation of the neutrinoless double beta decay 
effective Majorana mass in the TeV scale version of the 
type I seesaw model of neutrino mass generation 
\cite{Ibarra:2011xn}, etc.
}.
If CP invariance holds, we have
$\delta =0,\pi,2\pi$, the values 0 and $2\pi$ being physically 
indistinguishable, and
\cite{LW81} $\alpha_{21(31)} = k^{(')}\,\pi$, $k^{(')}=0,1,2,...$.

  The CP symmetry will not hold in the lepton sector 
if the Dirac and/or Majorana phases possess CP-nonconserving values.
If the Dirac phase $\delta$ 
has a CP-nonconserving value,
this will induce, as is well known,
CP violation effects in neutrino oscillations, i.e., 
a difference between the 3-flavour neutrino oscillation 
probabilities  $P(\nu_l \rightarrow \nu_{l'})$ and 
$P(\bar{\nu}_l \rightarrow \bar{\nu}_{l'})$, 
$l\neq l' =e,\mu,\tau$.  

  The flavour neutrino oscillation probabilities 
$P(\nu_l \rightarrow \nu_{l'})$ and 
$P(\bar{\nu}_l \rightarrow \bar{\nu}_{l'})$, 
$l,l' =e,\mu,\tau$, do not depend on the 
Majorana phases \cite{BHP80,Lang87}.
The Majorana phases can play important role in 
processes which are characteristic for Majorana neutrinos, 
in which the total lepton charge $L$ changes by two units, 
like neutrinoless double beta ($\betabeta$-) decay 
$(A,Z) \rightarrow (A,Z+2) + e^- + e^-$ (see, e.g., 
\cite{BiPet87,BPP1,WRodej10}), etc. 
The rates of the processes of emission of two different 
Majorana neutrinos, an example of which is the radiative 
emission of neutrino pair in atomic physics \cite{Yoshimura:2006nd}, 
depend in the threshold region on the Majorana 
phases \cite{Dinh:2012qb}.
The phases $\alpha_{21,31}$ 
can affect significantly the predictions for 
the rates of the lepton flavour violating (LFV) 
decays $\mu \rightarrow e + \gamma$,
$\tau \rightarrow \mu + \gamma$, etc.
in a large class of supersymmetric theories
incorporating the see-saw mechanism \cite{PPY03}. 

   Most importantly, the Dirac phases $\delta$ 
 and/or the Majorana phases 
 $\alpha_{21,31}$ in the PMNS neutrino mixing matrix can provide 
 the CP violation necessary for the generation 
 of the observed baryon asymmetry of the Universe 
 \cite{Pascoli:2006ie}.

 The existing neutrino oscillation data 
allow us to determine the two neutrino mass squared 
differences, $\Delta m^2_{21}$ and $|\Delta m^2_{31(32)}|$, 
and the three angles $\theta_{12}$,
$\theta_{23}$ and $\theta_{13}$, which 
drive the neutrino oscillations 
observed in the experiments 
with solar, atmospheric, reactor and accelerator 
neutrinos (see, e.g., \cite{PDG2014}) 
with a relatively good precision 
\cite{Capozzi:2013csa,GonzalezGarcia:2012sz}. 
The best fit values and the 3$\sigma$ allowed ranges of  
the three neutrino mixing parameters which are relevant for
our further discussion,
$\sin^2\theta_{12}$, $\sin^2\theta_{23}$ and $\sin^2\theta_{13}$,  
found in the global analysis in ref. \cite{Capozzi:2013csa} read:
\begin{eqnarray}
\label{th12values}
(\sin^2 \theta_{12})_{\rm BF} = 0.308,~~~~
 0.259 \leq \sin^2 \theta_{12} \leq 0.359\,,\\ [0.30cm]
\label{th23values}
(\sin^2\theta_{23})_{\rm BF} = 0.425~(0.437)\,,~~~~ 
 0.357(0.363) \leq \sin^2\theta_{23} \leq 0.641(0.659)\,,\\[0.30cm]
\label{th13values}
(\sin^2\theta_{13})_{\rm BF} = 0.0234~(0.0239)\,,~~~~ 
0.0177(0.0178) \leq \sin^2\theta_{13} \leq 0.0297(0.0300)\,,
\end{eqnarray}
%
where the value (the value in brackets) 
corresponds to  $\Delta m^2_{31(32)}>0$ ($\Delta m^2_{31(32)} <0$).
There are also hints from the data about the value of 
the Dirac phase
\footnote{Using the most recent T2K data on 
$\nu_{\mu}\rightarrow \nu_e$ oscillations, 
the T2K collaboration finds 
for $\delta = 0$, $\sin^2\theta_{23} = 0.5$ and 
$|\Delta m^2_{31(32)}| = 2.4\times 10^{-3}{\rm eV^2}$,
in the case of $\Delta m^2_{31(32)}>0$ 
($\Delta m^2_{31(32)} < 0$) \cite{T2K1113th13}: 
$\sin^22\theta_{13} = 0.140^{+0.038}_{-0.032}$ 
($0.170^{+0.045}_{-0.037}$). Thus, the best fit value of 
$\sin^22\theta_{13}$ thus found in the T2K experiment 
is approximately  by a factor of 1.6 (1.9) 
bigger than that measured in the Daya Bay 
experiment \cite{DBay1013th13}: 
$\sin^22\theta_{13} = 0.090^{+0.008}_{-0.009}$. 
The compatibility of 
the results of the two experiments on $\sin^22\theta_{13}$ 
requires, in particular, that  $\delta \neq 0$ 
(and/or $\sin^2\theta_{23} \neq 0.5$), 
which leads to the hints under discussion 
about the possible value of $\delta$ in the 
global analyses of the neutrino oscillation data.
}
$\delta$. In both 
analyses \cite{Capozzi:2013csa,GonzalezGarcia:2012sz}
the authors find that the best fit value of 
$\delta \cong 3\pi/2$. 
The CP conserving values $\delta = 0$ and $\pi$ 
($\delta = 0)$ are 
disfavoured at $1.6\sigma$ to  $2.0\sigma$ 
(at $2.0\sigma$) for  $\Delta m^2_{31(32)}>0$
($\Delta m^2_{31(32)}<0$). 
In the case of $\Delta m^2_{31(32)}<0$, the value 
$\delta = \pi$ is statistically $1\sigma$ away from 
the best fit value $\delta \cong 3\pi/2$ 
(see, e.g., Fig. 3 in ref.~\cite{Capozzi:2013csa}).

 The theoretical predictions for the values of the 
CPV phases in the neutrino mixing matrix depend on the 
approach and the type of symmetries 
one uses in the attempts to understand 
the pattern of neutrino mixing 
(see, e.g., \cite{Marzocca:2013cr,Girardi:2013sza,Luhn:2013lkn,Shimizu:2014ria}
and references quoted therein).
In the case of the Dirac phase $\delta$, the predictions 
vary considerably: they include the 
values 0, $\pi/2$, $\pi$, 3$\pi/2$, but not only; 
in certain cases  0, $\pi/2$, $\pi$ and 3$\pi/2$
are approximate values, the exact predictions being 
slightly different from these values. 
Obviously, a sufficiently precise measurement of $\delta$ 
will serve as an additional very useful constraint 
for identifying the approaches and/or the symmetries, 
if any, at the origin of the observed pattern of 
neutrino mixing.  Understanding the origin of the patterns of neutrino
masses and mixing, emerging from the neutrino oscillation,
$^3H$ $\beta-$decay, cosmological, etc.\ data is one of the most
challenging problems in neutrino physics.
It is part of the more general fundamental problem
in particle physics of understanding the origins of
flavour, i.e., of the patterns
of the quark, charged lepton and neutrino masses
and of the quark and lepton mixing.

 Using the fact that the neutrino mixing matrix 
$U = U^\dagger_{e}U_{\nu}$, where $U_{e}$ and $U_{\nu}$ 
result from the diagonalisation of the charged lepton 
and neutrino mass matrices, and assuming that 
$U_{\nu}$  has a i) tri-bimaximal (TBM) form \cite{TBM}, 
ii) bimaximal (BM) form \cite{SPPD82,BM}, 
or else iii) corresponds to the conservation of the 
lepton charge \cite{SPPD82} $L' = L_e - L_\mu - L_{\tau}$ (LC), 
that the requisite perturbative corrections 
to the TBM and BM (LC) mixing angles are provided by 
the matrix $U_e$, and that $U_e$ has a minimal 
form in terms of angles and phases it contains 
that can provide the corrections to 
$U_{\nu}$ so that the angles $\theta_{13}$, $\theta_{23}$
and $\theta_{12}$ in the PMNS matrix 
have values compatible with the current data, 
we have obtained  in \cite{Marzocca:2013cr}  
predictions for the Dirac phase $\delta$ 
present in the PMNS matrix $U$~
\footnote{The predictions for the Dirac phase $\delta$   
were obtained  in \cite{Marzocca:2013cr} 
using the framework which  was developed 
in \cite{GTani02,FPR04,SPWR04,Romanino:2004ww} 
for understanding the specific features of the 
neutrino mixing and in various versions 
was further exploited by many authors 
(see, e.g., \cite{Hochmuth:2007wq,Marzocca:2011dh,Alta,King2005,Antusch:2005kw,
 Antusch:2007rk,Antusch:2011qg,Chen:2009gf,Duarah:2012bm,Chao:2011sp}).
}.
An important requirement is that
the corrections due to the matrix $U_e$ should allow 
sizable deviations of the angle $\theta_{23}$ from the 
BM and TBM  value $\pm \pi/4$.
These requirements imply that 
$U_e$ should be a product of two rotations 
in the 12 and 23 planes, $R_{12}(\theta^e_{12})$ 
and $R_{23}(\theta^e_{23})$, and a diagonal 
phase matrix which contains, in general, 
two physical CP violation phases. 
In the case of ``standard'' ordering
with $U_e \propto R_{23}(\theta^e_{23}) R_{12}(\theta^e_{12})$, 
which we are going to consider and 
which is related to the 
hierarchy of the charged lepton masses, 
$m_e^2 \ll m^2_{\mu} \ll m^2_{\tau}$,  
and is a common feature of the overwhelming majority 
of the existing models of the charged lepton 
(and neutrino) masses and the associated mixing, 
$\cos\delta$ was shown to satisfy in the cases of the 
TBM and BM (LC) forms of $U_{\nu}$ 
a new sum rule \cite{Marzocca:2013cr} by which 
it is expressed in terms of the three angles 
$\theta_{12}$, $\theta_{23}$ and $\theta_{13}$ 
of the PMNS matrix. 
For the current best fit values of 
$\sin^2\theta_{12}$, $\sin^2\theta_{23}$ and $\sin^2\theta_{13}$, 
the following predictions for $\delta$ 
were obtained for the two forms of $U_{\nu}$  
\cite{Marzocca:2013cr}:
i) $\delta \cong \pi$ in the 
BM (or LC) case, ii) $\delta \cong 3\pi/2$ or $\pi/2$ 
in the TBM case
\footnote{More precisely, the predicted values of $\delta$ 
in the TBM case are  $\delta \cong 266^\circ$ or $94^\circ$.
}, 
the CP conserving values 
$\delta = 0, \pi, 2\pi$ being excluded in the TBM case 
at more than $5\sigma$. 
A model based on the $T^\prime$ flavour symmetry 
leading to the TBM form of $U_{\nu}$, in which the 
conditions of the general phenomenological approach 
followed in \cite{Marzocca:2013cr} 
are realised and thus which predicts, 
in particular,  $\delta \cong 3\pi/2$ or $\pi/2$, 
was constructed in \cite{Girardi:2013sza}.

  In the present article we first generalise in Section 3 
the analytic results for the sum rule 
involving the cosine of the Dirac phase $\delta$,  
obtained in \cite{Marzocca:2013cr} 
for the specific BM (LC) and TBM values 
$\pi/4$ and $\sin^{-1}(1/\sqrt{3})$ 
of the angle $\theta^{\nu}_{12}$ in the matrix 
$U_{\nu}$, to the case of arbitrary 
fixed value of $\theta^{\nu}_{12}$. 
This allows us to obtain {\it new} predictions 
for the phase $\delta$ and the 
$J_{\rm CP}$ factor, which controls the magnitude 
of CP violation effects due to $\delta$ 
in neutrino oscillations,
in the cases 
of i) golden ratio type A (GRA) mixing 
\cite{Everett:2008et,GRAM}
with $\sin^2\theta^{\nu}_{12} = (2 + r)^{-1} \cong 0.276$, 
$r$ being the golden ratio, 
$r = (1 +\sqrt{5})/2$, 
ii) golden ratio type B (GRB) mixing  
\cite{GRBM} with  
$\sin^2\theta^{\nu}_{12} = (3 - r)/4 \cong 0.345$,
and iii) hexagonal (HG) mixing 
\cite{HGM} in which $\theta^{\nu}_{12} = \pi/6$. 
As like the TBM and BM forms of $U_{\nu}$, the 
GRA form can be obtained from 
discrete family symmetry in the lepton sector, 
while the GRB and HG forms 
are considered on general phenomenological 
grounds (see, e.g., the reviews 
\cite{King:2013eh,Alta:2010ab,Tani:2010cd} and 
\cite{GRBM,HGM}).
In section 3 we derive also analytic expression for the 
correction in the new sum rule for $\cos\delta$ 
due to the possible presence in  $U_{e}$ 
of the 13 rotation matrix $R_{13}(\theta^e_{13})$ 
with angle $\theta^e_{13} \ll 1$ and determine 
the conditions under which this correction is 
sub-dominant.
 In Section 4 we show that the approximate 
sum rule for $\delta$ proposed in \cite{Antusch:2005kw}
can be obtained in the leading order approximation 
from the ``exact'' sum rule for $\cos\delta$ 
derived in Section 3. We compare the 
predictions for $\delta$ in the cases of 
the TBM, BM (LC), GRA, GRB and HG forms of 
of the matrix  $U_{\nu}$, obtained 
using the ``exact'' and the leading 
order sum rules and determine the origin 
of the difference in the predictions.
 We next analyse in Section 5 the possibility to 
obtain predictions for the values of the Majorana phases 
in the PMNS matrix, $\alpha_{21}$ and $\alpha_{31}$,
using the same approach which allowed 
us to get predictions for the Dirac phase $\delta$.   
For the  TBM, BM (LC), GRA, GRB and HG
forms of $U_{\nu}$ considered by us, 
we obtain analytic expressions 
for the contribution to the phases 
$\alpha_{21}$ and $\alpha_{31}$, 
generated by the CPV phases 
which serve in the approach 
employed as a ``source'' for the Dirac 
phase $\delta$ and which are 
present in the PMNS matrix due to 
the non-trivial form of the charged lepton 
``correction'' matrix  $U_e$.
We determine the cases when the 
phases $\alpha_{21,31}$
can be predicted and give example 
of prediction of their values.
We show in Section 6 that the results obtained 
on the Majorana phases for the different 
symmetry forms of the matrix  $U_{\nu}$,  
can lead, in particular, 
to specific predictions 
for the $\betabeta$-decay effective Majorana mass 
in the physically important cases of 
neutrino mass spectrum with inverted 
ordering or of quasi-degenerate type. 
The results of the present study 
are summarised in Section 7.

\vspace{-0.2cm}
%
\section{The Framework}
\label{Sec:GenSetup}
%

  In what follows we consider 
3-neutrino mixing of the three left-handed (LH) 
flavour neutrinos and antineutrinos, $\nu_l$ and 
$\bar{\nu}_l$, $l=e,\mu,\tau$.
The neutrino mixing matrix in this case receives 
contributions from the diagonalisation of the charged lepton 
and neutrino Majorana mass terms.  
Taking into account the contributions 
from the charged lepton and neutrino sectors, 
the PMNS neutrino mixing matrix can be written as \cite{FPR04}:
\begin{equation}
U_{\text{PMNS}}= U_e^{\dagger}\, U_{\nu} = 
(\tilde{U}_{e})^\dagger\, \Psi \tilde{U}_{\nu} \, Q_0\,.
\label{Ugeneral}
\end{equation}
%
Here  $U_e$ and $U_{\nu}$ 
are $3\times 3$ unitary matrices 
originating from the diagonalisation 
respectively of the charged lepton 
\footnote{For charged lepton mass term written 
in the left-right convention, the matrix $U_e$ 
diagonalises the hermitian matrix $M_E M^\dagger_E$, 
$U^\dagger_e M_E M^\dagger_E U_e = {\rm diag}(m^2_e,m^2_{\mu},m^2_{\tau})$, 
$M_E$ being the charged lepton mass matrix.
}
and neutrino mass matrices, 
$\tilde{U}_e$ and $\tilde{U}_\nu$ 
are CKM-like $3\times 3$ unitary matrices 
and  $\Psi$ and $Q_0$ are diagonal phase matrices 
each containing in the general case two physical CPV phases, 
\begin{equation} 
\Psi =
{\rm diag} \left(1,\text{e}^{-\ci \psi}, \text{e}^{-\ci \omega} \right)\,,~~
Q_0 = {\rm diag} \left(1,\text{e}^{\ci \frac{\xi_{21}}{2}}, 
\text{e}^{\ci \frac{\xi_{31}}{2}} \right)\,.
\label{PsieQ0}
\end{equation}
%
The phase matrix $Q_0$ contributes to the Majorana 
phases in the PMNS matrix and can appear in 
eq. (\ref{Ugeneral}) as a result of the diagonalisation of 
the neutrino Majorana mass term, while 
$\Psi$ can originate from the charged lepton sector 
($U_e^{\dagger} = (\tilde{U}_{e})^\dagger\, \Psi$),
or from the neutrino sector  
($U_{\nu} = \Psi \tilde{U}_{\nu} Q_0$), 
or  can receive contributions from both sectors.
  
 Following the results of the analysis performed in 
\cite{Marzocca:2013cr}, we will assume that 
the matrix $\tilde{U}_{e}$ is a product of two 
orthogonal matrices describing rotations 
in the 12 and 23 planes and that the 
two rotations in $\tilde{U}_e$ are
in the ``standard ordering''.
It proves convenient to adopt
for $\tilde{U}_e$ the notation used 
in \cite{Marzocca:2013cr}:
\begin{equation}
\tilde{U}_e =  R^{-1}_{23} \left( \theta^e_{23} \right)
R^{-1}_{12} \left( \theta^e_{12} \right)\,,
\label{tUe}
\end{equation}
%
where 
\begin{equation}
R_{12}\left( \theta^e_{12} \right) = \begin{pmatrix}
\cos \theta^e_{12} & \sin \theta^e_{12} & 0\\
- \sin \theta^e_{12} & \cos \theta^e_{12} & 0\\
0 & 0 & 1 \end{pmatrix} \;,
\quad
R_{23}\left( \theta^e_{23} \right) = \begin{pmatrix}
1 & 0 & 0\\
0 & \cos \theta^e_{23} & \sin \theta^e_{23} \\
0 & - \sin \theta^e_{23}  & \cos \theta^e_{23} \\
\end{pmatrix} \;,
\label{R1223}
\end{equation}
%
and $\theta^e_{12}$ and $\theta^e_{23}$
are two arbitrary (real) angles. 

The fact that  $\tilde{U}_{e}$ does not include 
the matrix $R_{13}(\theta^e_{13})$ describing rotation 
in the 13 plane, i.e.,  
that  $\theta^e_{13} \cong 0$,
follows from the requirement that $U_e$ has a  
 ``minimal'' form in terms of angles 
and phases it contains that can provide 
the requisite corrections to 
$U_{\nu}$, so that the mixing angles 
$\theta_{13}$, $\theta_{23}$
and $\theta_{12}$ in $U$
have values compatible with the current data, 
including the possibility of a sizable deviation 
of $\theta_{23}$ from $\pi/4$. 
As will be discussed briefly in Section 3, 
a nonzero $\theta^e_{13} \ltap 10^{-3}$  
generates a correction to $\cos\delta$ 
derived from the exact sum rule, which does not exceed 
11\% (4.9\%) in the TBM (GRB) cases and is even smaller 
in the other three cases of symmetry form 
of $\tilde{U}_{\nu}$ analysed in the 
present article. 
We note that $\theta^e_{13} \cong 0$ 
is a feature of many  theories and models 
of charged lepton mass generation 
(see, e.g., \cite{Girardi:2013sza,Antusch:2011qg,Chen:2009gf,Everett:2008et,King:2013eh,CarlMCC})
and was used in a large number of 
articles dedicated to the problem 
of understanding the origins of the observed pattern
of neutrino mixing 
(see, e.g., \cite{Shimizu:2014ria,GTani02,FPR04,Romanino:2004ww,Antusch:2005kw,Antusch:2007rk,Duarah:2012bm,Chao:2011sp,Kile:2014kya,Hall:2013yha}).
In large class of GUT inspired 
models of flavour, for instance, 
the matrix $U_e$ is directly related to 
the quark mixing matrix 
(see, e.g., \cite{Marzocca:2011dh,Antusch:2011qg,Chen:2009gf,King:2013eh,Alta:2010ab}). 
As a consequence, in this class of
models, in particular,
$\theta^e_{13}$ is negligibly small.

 We will assume further that 
the matrix $\tilde{U}_{\nu}$ has one of the 
following symmetry forms: 
TBM, BM, LC, GRA, GRB and HG. 
For all symmetry forms of interest, 
$\tilde{U}_{\nu}$ is also a product 
23 and 12 rotations in the plane: 
\begin{equation}
\tilde{U}_\nu =  R_{23} \left( \theta^\nu_{23} \right)
R_{12} \left( \theta^\nu_{12} \right)\,.
\label{Unu}
\end{equation}
%
In the case of the TBM, BM, GRA, GRB and HG
forms of $\tilde{U}_\nu$ we have  $\theta^\nu_{23} = -\,\pi/4$, 
while  $\theta^\nu_{12}$ takes the values 
$\sin^{-1}(1/\sqrt{3})$, $\pi/4$,  
$\sin^{-1}(1/\sqrt{2 + r})$, 
$\sin^{-1}(\sqrt{3 - r}/2)$, 
and $\pi/6$, respectively. 
Thus, the matrix $\tilde{U}_{\nu}$ corresponding 
to these cases has the form:
\begin{equation}
\tilde{U}_{\nu} = \begin{pmatrix}
\cos \theta^{\nu}_{12} & \sin \theta^{\nu}_{12} & 0\\
- \frac{\sin \theta^{\nu}_{12}}{\sqrt{2}} & 
\frac{\cos \theta^{\nu}_{12}}{\sqrt{2}} & 
- \frac{1}{\sqrt{2}}   \\
- \frac{\sin \theta^{\nu}_{12}}{\sqrt{2}}  & 
\frac{\cos \theta^{\nu}_{12}}{\sqrt{2}} & 
\frac{1}{\sqrt{2}} 
\end{pmatrix} \;,
\label{Unu1}
\end{equation}
%
where $\theta^{\nu}_{12}$ takes different 
fixed values for the different symmetry 
forms of $\Tilde{U}_{\nu}$.
In the case of the LC form of $\tilde{U}_{\nu}$ 
we have $\theta^{\nu}_{12}= \pi/4$, while 
$\theta^{\nu}_{23}$ can have an arbitrary 
fixed value.
Thus, if  $U_e = {\bf 1}$, ${\bf 1}$ 
being the unity $3\times 3$ matrix, 
we have:\\
i) $\theta_{13} = 0$ in all six cases 
of interest of $\tilde{U}_\nu$;\\
ii) $\theta_{23}= - \pi/4$, 
if $\tilde{U}_\nu$ has  
any of the forms  TBM, BM, GRA, GRB and HG, 
while  $\theta_{23}$ can have an 
arbitrary value if $U_\nu$ has the 
LC form;\\
iii) $\sin^2\theta_{12}= 0.5$ 
for the BM and LC forms of  $\tilde{U}_\nu$;
$\sin^2\theta_{12}= 1/3$ in the TBM case;
$\sin^2\theta_{12} \cong 0.276$ and 0.345 
for the GRA and GRB mixing and 
 $\sin^2\theta_{12}= 0.25$ for 
HG mixing.  
Thus, the matrix $U_e$ has to generate 
corrections \\
i) leading to $\theta_{13} \neq 0$ compatible 
with the observations in all six cases of 
$U_\nu$  considered;\\
ii) leading to the observed deviation of 
$\theta_{12}$ from a) $\pi/4$, b) from the two 
golden ratio values 
\footnote{The GRA and GRB values of 
$\sin^2\theta_{12} \cong 0.276$ and 0.345 
lie at the border of the $2\sigma$
allowed range of values of $\sin^2\theta_{12}$ 
obtained in the global analyses 
\cite{Capozzi:2013csa,GonzalezGarcia:2012sz}.} 
and c) from $\pi/6$,  
in the cases of a) BM and LC, 
b) GRA and GRB, and c) HG,  mixing;\\
iii) leading to the sizable
deviation of $\theta_{23}$ from $\pi/4$
for all cases considered except the LC one,
if it is confirmed by further data 
that $\sin^2\theta_{23} \cong 0.40 - 0.44$.
The minimal form of $U_e$
in terms of angles and phases it contains, 
which can produce the requisite corrections 
discussed above, is the one with 
$\tilde{U}_e$ given in eq. (\ref{tUe}). 
The presence of $R^{-1}_{12}(\theta^e_{12})$ 
in $\tilde{U}_e$ allows to correct  
the symmetry values of $\theta_{12}$ and 
$\theta_{13}$, while the presence of
$R^{-1}_{23}(\theta^e_{23})$ allows to have 
sizable deviations (bigger than 
$0.5\sin^2\theta_{13}$) of 
$\sin^2\theta_{23}$ from the symmetry 
value of 0.5.  
 
  In the approach adopted by us following 
\cite{Marzocca:2013cr} the PMNS neutrino mixing matrix 
has the form:
\begin{equation}
U_{\text{PMNS}}=  U_{e}^\dagger U_{\nu}
= R_{12}( \theta^e_{12})R_{23}(\theta^e_{23})  \Psi\,
 R_{23}(\theta^{\nu}_{23}) R_{12}(\theta^{\nu}_{12})\; Q_{0}\,.
\label{UPMNS2}
\end{equation}
%
where $\theta^{\nu}_{23} = -\,\pi/4$ and 
$\theta^{\nu}_{12}$ has a known value.
As a consequence, the three angles 
$\theta_{12}$, $\theta_{23}$ and $\theta_{13}$
and the Dirac CPV phase $\delta$ of the PMNS mixing matrix, 
eqs. (\ref{eq:VQ}) - (\ref{eq:Vpara}),
can be expressed as functions of the two real angles,
$\theta^e_{12}$ and $\theta^e_{23}$,
and the two phases, $\psi$ and $\omega$, of the 
phase matrix $\Psi$. The results will depend 
on the specific value of the angle 
$\theta^{\nu}_{12}$, i.e., on the assumed 
symmetry form of $\tilde{U}_{\nu}$.
We will discuss how the Majorana phases 
in the PMNS matrix, $\alpha_{21}$ and $\alpha_{31}$, 
are expressed in terms of these parameters later. 

 As was shown in \cite{Marzocca:2013cr}, 
the product of matrices $R_{23}(\theta^e_{23})  \Psi
 R_{23}(\theta^{\nu}_{23} = -\pi/4)$ in the expression 
(\ref{UPMNS2}) for $U_{\text{PMNS}}$  
can be rearranged as follows:
\begin{equation}
R_{23}( \theta^e_{23})\, \Psi \, R_{23}(\theta^{\nu}_{23}) =
P_1\, \Phi\, R_{23}(\hat\theta_{23})\,Q_1\,.
\label{Phi}
\end{equation}
%
Here the angle $\hat\theta_{23}$ is determined by 
\begin{equation}
\sin^2\hat\theta_{23} =
\frac{1}{2}\,\left (1 - 2\sin\theta^e_{23}\cos \theta^e_{23}\cos(\omega -\psi)\right )\,,
\label{th23hat}
\end{equation}
%
and
\begin{equation}
P_1={\rm diag}(1,1, \text{e}^{-\,\ci \alpha})\,,~
\Phi = {\rm diag}(1,\text{e}^{\ci \phi},1)\,,~
Q_1 = {\rm diag} \left(1,1, \text{e}^{\ci \beta} \right)\,,
\label{PPhitQ}
\end{equation}
%
where
\begin{equation}
\alpha = \gamma + \psi + \omega  \,,~~~~
\beta = \gamma - \phi\,,
\label{alphabeta}
\end{equation}
%
and
\begin{equation}
\gamma = \arg \left (\,-\text{e}^{ -\ci \psi}\cos \theta^e_{23}
+ \text{e}^{-\ci \omega}\sin\theta^e_{23}\right)\,,~~
\phi= \arg \left (\text{e}^{ -\ci \psi}\cos \theta^e_{23}
+ \text{e}^{-\ci \omega}\sin\theta^e_{23}\right)\,.
\label{gammaphi}
\end{equation}
%
The phase $\alpha$ in the matrix $P_1$ is unphysical.
The phase $\beta$ contributes
to the matrix of physical Majorana phases,
which now is equal to $\hat{Q} = Q_1\,Q_0$.
The PMNS matrix takes the form:
\begin{equation}
U_{\text{PMNS}}=
R_{12}(\theta^e_{12})\,\Phi(\phi)\, R_{23}(\hat\theta_{23})\,
R_{12}(\theta^{\nu}_{12})\,\hat{Q}\,,
\label{UPMNSthhat1}
\end{equation}
%
where $\theta^{\nu}_{12}$ has a fixed value 
which depends on the symmetry form of $\tilde{U}_\nu$ used.
Thus, the four observables $\theta_{12}$, $\theta_{23}$, $\theta_{13}$
and  $\delta$ are functions of three parameters
$\theta^e_{12}$, $\hat\theta_{23}$ and $\phi$.
As a consequence, the Dirac phase $\delta$ can be expressed
as a function of the three PMNS angles
$\theta_{12}$, $\theta_{23}$ and $\theta_{13}$,
leading to a new ``sum rule''
relating $\delta$ and $\theta_{12}$, $\theta_{23}$ and $\theta_{13}$
\cite{Marzocca:2013cr}. Using the measured values of
$\theta_{12}$, $\theta_{23}$ and $\theta_{13}$,
we have obtained in \cite{Marzocca:2013cr}
predictions for the values of $\delta$ and of
the rephasing invariant 
$J_{\text{CP}} = {\rm Im}(U_{e1}^* U_{\mu 3}^* U_{e3} U_{\mu 1})$,
which controls the magnitude of CP violating effects
in neutrino oscillations \cite{PKSP3nu88}, 
in the cases of the TBM, BM (LC) forms 
of $\tilde{U}_{\nu}$. Here we will first 
obtain predictions for $\delta$ and $J_{\text{CP}}$ 
in the cases of GRA, GRB and HG  forms of $\tilde{U}_{\nu}$.
After that we will analyse the possibility 
to obtain predictions for the Majorana phases 
in the PMNS matrix within the framework 
described above.

\vspace{-0.3cm}
%
\section{The Dirac Phase in the PMNS Matrix}
\label{Sec:GenSetup}
%

Using eq.~(\ref{UPMNSthhat1})
we get for the angles $\theta_{12}$, $\theta_{23}$ and $\theta_{13}$
of the standard parametrisation of $U_{\text{PMNS}}$
\cite{Marzocca:2013cr}:
\begin{align}
\label{s2th13}
    \sin \theta_{13} &= \left| U_{e3} \right| =
\sin \theta^e_{12} \sin \hat{\theta}_{23}, \\[0.30cm]
\label{s2th23}
\sin^2 \theta_{23} &=
\frac{\left| U_{\mu3} \right|^2}{1- \left| U_{e 3} \right|^2 } 
= \sin^2 \hat{\theta}_{23}\, 
\frac{\cos^2 \theta_{12}^e}{1 - \sin^2 \theta^e_{12} \sin^2 \hat{\theta}_{23} }
= \frac{\sin^2\hat\theta_{23} - \sin^2 \theta_{13}}
{1 - \sin^2\theta_{13}}\,, \\[0.30cm]
\sin^2 \theta_{12} &= 
\frac{\left| U_{e2} \right|^2}{1- \left| U_{e3} \right|^2 } =
\frac{\left| \sin \theta_{12}^\nu \cos \theta_{12}^e 
+ e^{i \phi} \cos \theta_{12}^\nu \cos \hat{\theta}_{23} 
\sin \theta_{12}^e \right|^2}
{1 - \sin^2 \theta^e_{12} \sin^2 \hat{\theta}_{23} }\,
\label{s2th12}
\end{align}
%
where eq. (\ref{s2th13})
was used in order to obtain the expression for
$\sin^2 \theta_{23}$ in terms of $\hat{\theta}_{23}$ and $\theta_{13}$.
Within the approach employed the expressions in 
eqs. (\ref{s2th13}) - (\ref{s2th12}) 
are exact. It follows from eqs. (\ref{s2th13}) and  (\ref{s2th23})
that the angle $\hat\theta_{23}$ differs little from
the angle $\theta_{23}$ 
and that $\sin^2 \theta^e_{12} \ll 1$: 
for, e.g., the best fit values of $\sin^2\theta_{13} = 0.0234$
and $\sin^2\theta_{23} \cong 0.425$ 
we have  $\sin^2\hat{\theta}_{23} \cong 0.438$
and $\sin \theta^e_{12}\cong 0.23$.

 We will derive next 
first a general expression for 
the cosine of the CPV phase $\phi$ 
in terms of the angles $\theta_{12}$, 
$\theta_{23}$, $\theta_{13}$ and $\theta^{\nu}_{12}$, 
then a relation between the phases $\phi$ 
and the Dirac phase $\delta$ of the standard 
parametrisation of the PMNS matrix, 
and finally an expression for $\delta$ in terms of 
of the angles 
$\theta_{12}$, $\theta_{23}$, 
$\theta_{13}$, $\theta^{\nu}_{12}$ and  
for an arbitrary fixed value of 
$\theta^{\nu}_{12}$.
This will allow us to obtain new predictions for 
$\delta$ in the cases of 
GRA, GRB and HG symmetry forms 
of the matrix $\tilde{U}_{\nu}$.

 From eq. (\ref{s2th12}) using 
eqs. (\ref{s2th13}) and  (\ref{s2th23}) we find:
\begin{equation} 
\cos\phi = 2\,\frac{
\sin^2\theta_{12}\,(1 - \cos^2\theta_{23}\cos^2\theta_{13}) - 
(\sin^2\theta_{23}\sin^2\theta^{\nu}_{12} +
\cos^2\theta_{23}\cos^2\theta^{\nu}_{12}\sin^2\theta_{13})}
{\sin2\theta^{\nu}_{12} \sin2\theta_{23}\,\sin\theta_{13}}\,.
\label{cphi}
\end{equation}
%
As it follows from eqs. (\ref{Phi}), (\ref{PPhitQ}) and (\ref{alphabeta}), 
the phase $\phi$ contributes to the Majorana phase $\alpha_{31}$, 
in particular,  
via the phase $\beta$. Thus, we will give next the values of 
$\cos\phi$ and $|\sin\phi|$ for the different symmetry forms of the matrix 
$\tilde{U}_{\nu}$ we are considering, TBM, BM (LC), GRA, GRB and HG
\footnote{Using the current data one can determine directly only 
$\cos\phi$ but not $\sin\phi$, and therefore the sign of $\sin\phi$ 
is undetermined. The measurement of $\sin\delta$ 
will allow to determine $\sin\phi$ as well.
}. 
These values will be relevant in the discussion of the Majorana phases 
determination. Using the best fit values of the neutrino mixing parameters 
$\sin^2\theta_{12}$, $\sin^2\theta_{23}$ and $\sin^2\theta_{13}$ 
quoted in eqs. (\ref{th12values}) - (\ref{th13values}),  
for $\Delta m^2_{31} > 0$ 
and the specific value of $\theta^{\nu}_{12}$ characterising 
a given case of $\tilde{U}_\nu$, we get:
\begin{align}
\label{cosphTBM}
{\rm TBM: } \qquad& \cos\phi \cong -\, 0.219\,,~~~
 |\sin\phi| \cong  0.976\,,&\\[0.25cm]
\label{osphGRA}
{\rm GRA: } \qquad& \cos\phi \cong +\, 0.116\,,~~~
|\sin\phi| \cong  0.993\,,& \\[0.25cm]
\label{cosphGRB}
{\rm GRB: } \qquad& 
\cos\phi \cong -\, 0.286\,,~~~
|\sin\phi| \cong  0.958\,,& \\[0.25cm]
\label{cosphHG}
{\rm HG: }  \qquad& 
\cos\phi \cong  +\, 0.286\,,~~~
|\sin\phi| \cong  0.958\,.
\end{align}
%
The same procedure leads in the BM (LC) 
case to the unphysical value of 
$\cos\phi \cong - 1.13$. 
This reflects the fact that
the scheme under 
discussion with BM (LC) form of the matrix 
$\tilde{U}_{\nu}$ does not provide a good 
description of the current data on 
$\theta_{12}$, $\theta_{23}$ and $\theta_{13}$  
\cite{Marzocca:2013cr}.
Thus, we will calculate $\cos\phi$ 
using  the best values of $\sin^2\theta_{12} = 0.32$, 
$\sin^2\theta_{23} = 0.41~(0.42)$ and 
$\sin\theta_{13} = 0.158$, determined 
for $\Delta m^2_{31} > 0$ 
($\Delta m^2_{31} < 0$) in the statistical 
analysis performed in \cite{Marzocca:2013cr}. 
For these values of $\sin^2\theta_{12}$, 
$\sin^2\theta_{23}$ and $\sin\theta_{13}$ 
in the case of $\Delta m^2_{31} > 0$ we get:
\begin{align}
\label{cosphBM}
 {\rm BM~(LC): }  \qquad& 
\cos\phi \cong -\, 0.981\,,~~~ |\sin\phi| \cong 0.193\,,
\end{align}
%
We do not give the results on $\cos\phi$ for 
$\Delta m^2_{31} < 0$ since they differ little 
from those shown.

Comparing the imaginary and real parts of
$U_{e1}^* U_{\mu 3}^* U_{e3} U_{\mu 1}$, obtained
using eq.~(\ref{UPMNSthhat1})
and the standard parametrisation of $U_{\text{PMNS}}$,
one gets the following relation between 
$\phi$ and $\delta$:
\begin{align}
\label{sindsinphi}
\sin\delta =
&\; -\, \frac{\sin2\theta^{\nu}_{12}}{\sin2\theta_{12}}\,\sin\phi\,,
\\[0.30cm]
\cos\delta =&\; \frac{\sin2\theta^{\nu}_{12}}{\sin2\theta_{12}}\,\cos\phi\,
\left (-1 + \frac{2\sin^2\theta_{23}}
{\sin^2\theta_{23}\cos^2\theta_{13} + \sin^2\theta_{13}}\,\right )
+\,\frac{\cos2\theta^{\nu}_{12}}{\sin2\theta_{12}}\,
\frac{\sin2\theta_{23}\, \sin\theta_{13}}
{\sin^2\theta_{23}\cos^2\theta_{13} + \sin^2\theta_{13}}\,.
\label{cosdcosphi}
\end{align}
%
Within the scheme considered the results quoted above, including those
for $\sin\delta$ and $\cos\delta$, are exact 
and are valid for arbitrary fixed $\theta^{\nu}_{12}$.
As can be shown, in particular, we have:
$\sin^2\delta + \cos^2\delta = 1$. In Section 5
we will derive an exact relation between 
the CPV phases $\delta$ and $\phi$ 
(see eq. (\ref{dphibeta})).

  Substituting the expression (\ref{cphi}) for 
$\cos\phi$ in eqs. (\ref{sindsinphi}) and (\ref{cosdcosphi}),
we get a general expressions for $\sin\delta$ and $\cos\delta$
in terms of $\theta_{12}$, $\theta_{23}$, $\theta_{13}$ and 
$\theta^{\nu}_{12}$. We give below the result for $\cos\delta$:
\begin{equation}
\cos\delta =  \frac{\tan\theta_{23}}{\sin2\theta_{12}\sin\theta_{13}}\,
\left [\cos2\theta^{\nu}_{12} + 
\left (\sin^2\theta_{12} - \cos^2\theta^{\nu}_{12} \right )\,
 \left (1 - \cot^2\theta_{23}\,\sin^2\theta_{13}\right )\right ]\,.
\label{cosdthnu}
\end{equation}
%
For $\theta^{\nu}_{12} = \pi/4$ and $\theta^{\nu}_{12}= \sin^{-1}(1/\sqrt{3})$ 
the expression (\ref{cosdthnu}) for $\cos\delta$ we have derived 
reduces to those found in \cite{Marzocca:2013cr} 
in the BM (LC) and TBM cases, respectively.

From eq. (\ref{cosdthnu}) we find in the cases of 
TBM and BM (LC) forms of 
$\tilde{U}_{\nu}$
\footnote{There is a small difference between the
values of $\cos\delta$ and $\delta$ obtained 
for $\Delta m^2_{31} > 0$ and $\Delta m^2_{31} < 0$. 
We report here the values corresponding 
to  $\Delta m^2_{31} > 0$.
}:
\begin{align}
\label{cosdTBM}
{\rm TBM: } \qquad& \cos\delta = -\, 0.0851\,,~~~
\delta = 265.1^\circ~{\rm or}~94.9^\circ\,,&\\[0.30cm]
\label{cosdBM}
{\rm BM~(LC): }  \qquad& \cos\delta = -\, 0.978,,~~~
\delta = 191.9^\circ~{\rm or}~168.1^\circ\,.
\end{align}
%
The value of $\cos\delta$ corresponds 
in the TBM case to the 
 best fit values 
 of $\sin^2\theta_{ij}$ given 
in eqs. (\ref{th12values}) - (\ref{th13values});
in the BM (LC) case 
it is obtained for 
\cite{Marzocca:2013cr}  
$\sin^2\theta_{12}$ = 0.32, 
$\sin^2\theta_{23}$ = 0.41 and 
$\sin\theta_{13}$ = 0.158.

 For the new cases considered by us 
we get using the best fit values of $\sin^2\theta_{ij}$
quoted in eqs. (\ref{th12values}) - (\ref{th13values}) 
for $\Delta m^2_{31} > 0$ ( $\Delta m^2_{31} < 0$) and the 
value of $\theta^{\nu}_{12}$ characterising a given case:
\begin{align}
\label{cosdGRA}
{\rm GRA: }\qquad& \cos\delta \cong 0.273~(0.274)\,,~~
\delta \cong 285.8^\circ~(285.9^\circ)~{\rm or}~
74.2^\circ~(74.1^\circ)\,,& \\[0.30cm]
\label{cosdGRB}
{\rm GRB: }\qquad& \cos\delta \cong -\, 0.161~(-0.165)\,,~~
\delta \cong 260.7^\circ~(260.5^\circ)~{\rm or}~
99.3^\circ~(99.5^\circ)\,,& \\[0.30cm]
\label{cosdHG}
{\rm HG: }\qquad& \cos\delta \cong  0.438~(0.442)\,,~~
\delta \cong 296.0^\circ~(296.2^\circ)~{\rm or}~
64.0^\circ~(63.8^\circ)\,. 
\end{align}
%
It follows from the results derived 
and quoted above that, in general,  
the predicted values of $\cos\delta$ and $\delta$ 
vary significantly 
with the assumed symmetry form of the 
matrix $\tilde{U}_\nu$.
One exception are the predictions of $\delta$ 
in the cases of TBM and GRB forms of $\tilde{U}_{\nu}$: 
they differ only by approximately $5^\circ$.
We note also that, except for the BM (LC) case, 
the values of $\cos\delta$ and $\cos\phi$ 
differ significantly for a given assumed 
form of the symmetry mixing, TBM, GRA, etc. 

   If we consider the indications obtained in 
\cite{Capozzi:2013csa,GonzalezGarcia:2012sz} 
that $\delta \cong 3\pi/2$, only
the case of BM (LC) mixing is weakly disfavoured for $\Delta m^2_{31} > 0$ at 
approximately $1.4\sigma$, while for $\Delta m^2_{31} < 0$ 
all cases of the form of $\tilde{U}_\nu$
considered by us are statistically  
compatible with the results on $\delta$ 
found in \cite{Capozzi:2013csa,GonzalezGarcia:2012sz} 
(see, e.g., Fig. 3 in \cite{Capozzi:2013csa}).

  As was mentioned in Section 2, 
a nonzero $|\sin\theta^e_{13}| \ll 1$, 
$\theta^e_{13}$ being the angle of rotation in the 
13 plane, generates a correction to the value of $\cos\delta$ 
derived from the exact sum rule. 
In this case we have:
$\cos\delta(\theta^e_{13}) = \cos\delta - \Delta(\cos\delta)$, 
where $\cos\delta$ is the value obtained 
from the exact sum rule and $\Delta(\cos\delta)$
is the correction due to  $|\sin\theta^e_{13}| \neq 0$.
As can be shown using the parametrisation 
$\tilde{U}_e =  R^{-1}_{23}(\theta^e_{23}) R^{-1}_{13}(\theta^e_{13})
R^{-1}_{12}(\theta^e_{12})$, 
to leading order in $|\sin\theta^e_{13}| \ll 1$ 
(i.e., neglecting terms of order of, or smaller than, 
$\sin^2\theta^e_{13}$, $\sin\theta^e_{13}\sin\theta_{13}$)
we have:
\begin{equation}
\Delta(\cos\delta) \cong 
\frac{\sin\theta^e_{13}}{\sin\theta_{13}}\, 
\frac{\cos\kappa}{\sin\hat\theta_{23}}\,
\tan\theta^{\nu}_{12}\,\cot\theta_{12}\, \tan\theta_{23}\,,
\label{th13e}
\end{equation}
%
where $\kappa = {\rm arg}(c^e_{23}e^{-\ci \omega} - s^e_{23}e^{-\ci\psi})$.
The result (\ref{th13e}) for $\Delta(\cos\delta)$ can be derived 
by taking into account, in particular,  
that $|\sin\theta^e_{13}|\ll 1$ 
and that in the approximation employed by us 
$\cos\delta(\theta^e_{13}) \sin\theta_{13} 
\cong \cos\delta\, \sin\theta_{13}$.
It is not difficult to convince oneself that 
for the best fit values of the neutrino mixing 
parameters and the symmetry forms of $\tilde{U}_{\nu}$ 
considered, the correction satisfies the inequality:
$|\Delta(\cos\delta)| \ltap C\,|\sin\theta^e_{13}|$,
where the constant $C = 9.0$, 12.7, 7.9, 9.2, and 7.3 
for the TBM, BM, GRA, GRB and HG forms of  $\tilde{U}_{\nu}$, 
respectively. Thus, for  $|\sin\theta^e_{13}| \ltap 10^{-3}$,
the correction $|\Delta(\cos\delta)|$ 
to the exact sum rule result for $\cos\delta$
does not exceed 11\% (4.9\%) in the case of the TBM (GRB) 
form and is even smaller 
for the BM, GRA and HG  forms of  $\tilde{U}_{\nu}$.
In what follows we concentrate on the case 
of negligibly small $\sin\theta^e_{13} \cong 0$. 

    The fact that the value of the Dirac CPV phase
$\delta$ is determined (up to an ambiguity of
the sign of $\sin\delta$) by the values of the
three mixing angles  $\theta_{12}$, $\theta_{23}$
and $\theta_{13}$ of the PMNS matrix 
and the value of $\theta^{\nu}_{12}$ of 
the matrix $\tilde{U}_\nu$, eq. (\ref{Unu1}), 
is the most striking prediction of the
model considered. This result implies 
also that in the scheme
under discussion, the rephasing invariant $J_{\text{CP}}$
associated with the Dirac phase $\delta$, 
which determines the magnitude
of CP violation effects in neutrino oscillations 
\cite{PKSP3nu88} and in the standard parametrisation of 
the PMNS matrix has the well known form,
\be
J_{CP} = \text{Im} \left\{ U_{e1}^* U_{\mu3}^* U_{e3} U_{\mu1} \right\} = 
\frac{1}{8} \sin \delta \sin 2 \theta_{13} \sin 2 \theta_{23} 
\sin 2 \theta_{12} \cos \theta_{13}\,,
\label{JCP}
\ee
%
is also a function of the three angles
$\theta_{12}$, $\theta_{23}$ and $\theta_{13}$
of the PMNS matrix and of $\theta^{\nu}_{12}$:
\begin{equation}
J_{\text{CP}} = J_{\text{CP}}(\theta_{12},\theta_{23},\theta_{13},
\delta(\theta_{12},\theta_{23},\theta_{13},\theta^{\nu}_{12})) =
J_{\text{CP}}(\theta_{12},\theta_{23},\theta_{13},\theta^{\nu}_{12})\,.
\label{JCPNO}
\end{equation}
%
This allows us to obtain predictions for the range of
possible values of $J_{\text{CP}}$ in the cases of 
different symmetry forms of $\tilde{U}_\nu$, which are 
specified by the value of $\theta^{\nu}_{12}$, 
using the current data on $\sin^2\theta_{12}$, 
$\sin^2\theta_{23}$ and $\sin^2\theta_{13}$. 
Using the best fit values 
of the neutrino mixing angles,
we have for  $\Delta m^2_{31} > 0$ 
($\Delta m^2_{31} < 0$):
\begin{align}
\label{JCPTBM}
{\rm TBM: }   \qquad& J_{\rm CP} \cong \mp\, 0.034\,,&\\[0.25cm]
\label{JCPBM}
{\rm BM~(LC): }  \qquad& J_{\rm CP} \cong \mp\,0.008~(\mp 0.003)\,,&\\[0.25cm]
\label{JCPGRA}
{\rm GRA: }   \qquad&  J_{\rm CP} \cong \mp 0.0328~(\mp 0.0332)\,,& \\[0.25cm]
\label{JCPGRB}
{\rm GRB: }   \qquad& J_{\rm CP} \cong \mp 0.0336~(\mp 0.0341)\,,& \\[0.25cm]
\label{JCPHG} 
{\rm HG: }   \qquad&  J_{\rm CP} \cong \mp 0.0306~(\mp 0.0310)\,. 
\end{align}
%
where the results in the TBM and BM cases were 
obtained in \cite{Marzocca:2013cr} 
\footnote{The statistical analyses performed in 
\cite{Marzocca:2013cr} showed, in particular, that, given the indication 
for $\delta \cong 3\pi/2$ found in the global analyses of the
current neutrino oscillation data, 
in the TBM case  the value of  $J_{\rm CP} \cong +\, 0.034$  
is statistically disfavoured with respect to the value 
$J_{\rm CP} \cong -\, 0.034$.}. 
It follows from eqs. (\ref{JCPTBM}) - (\ref{JCPHG})
that,  apart from the BM (LC) case, 
the $|J_{\rm CP}|$ factor has rather similar values 
in the TBM, GRA, GRB and HG mixing cases. 
As our results show, distinguishing between these cases 
requires a measurement of $\cos\delta$ or a very 
high precision measurement of $|J_{\rm CP}|$. 

\vspace{-0.3cm}
%
\section{The Case of  $|\sin\theta^e_{23}| \ll 1$}
%
%
\subsection{Negligible $\theta^e_{23}$}
%
%
 The case of negligible $\theta^e_{23} \cong 0$ was analysed by many authors 
(see, e.g., \cite{GTani02,FPR04,SPWR04,Romanino:2004ww,Hochmuth:2007wq,
 Marzocca:2011dh,Alta,Antusch:2005kw} 
as well as \cite{Shimizu:2014ria}).
It corresponds to a large number of theories and models of 
charged lepton and neutrino mass generation 
(see, e.g., 
\cite{Marzocca:2011dh,Alta,Antusch:2011qg,Chen:2009gf,Chao:2011sp,King:2013eh}).
  In the limit of negligibly small $\theta^e_{23}$ we find from 
eqs. (\ref{th23hat}), (\ref{alphabeta}) and (\ref{gammaphi}):
\begin{equation}
\sin^2\hat\theta_{23} =\frac{1}{2}\,,~~~\gamma= -\psi + \pi\,,~~~
\phi = - \psi\,,~~~ \beta = \gamma - \phi = \pi\,.
\label{th23hatphases}
\end{equation}
%
The phase $\omega$ is unphysical.
All results obtained in the previous section 
are valid also in the case of negligibly 
small $\theta^e_{23}$: one has to set 
$\sin^2\hat\theta_{23} = 0.5$ 
in the expressions derived for arbitrary 
$\sin^2\hat\theta_{23}$
in the preceding Section. From 
eqs. (\ref{s2th13}) -(\ref{s2th12}), using 
the fact that $\sin^2\hat\theta_{23} = 0.5$,
we get the well known results for 
$\sin \theta_{13}$ and $\sin^2 \theta_{23}$,
\begin{align}
\label{s2th13se230}
    \sin \theta_{13} & = \frac{1}{\sqrt{2}}
\sin \theta^e_{12}\,,~~\sin^2 \theta_{23} =
 \frac{1 - 2\sin^2 \theta_{13}}{2(1 - \sin^2\theta_{13})}
\cong \frac{1}{2}\,\left (1 -  \sin^2\theta_{13}\right )\,,
\end{align}
%
and the following new exact expression for $\sin^2\theta_{12}$:
\begin{align}
\sin^2\theta_{12} & =
\sin^2\theta_{12}^\nu + 
\cos2\theta_{12}^\nu \,\frac{\sin^2\theta_{13}}{1-\sin^2\theta_{13}}
+ \sin 2\theta^\nu_{12}\,  \sin\theta_{13}\,\cos\phi \, 
 \frac{(1 - 2\sin^2\theta_{13})^{\frac{1}{2}}}
{1 - \sin^2\theta_{13} }\,.
\label{s2th12se230phi}
\end{align}
%
In the case of $\theta^e_{23} = 0$, as is well known, 
$\sin^2\theta_{23}$ can deviate only 
by $0.5\sin^2\theta_{13}$ from 0.5.
Let us emphasise that the exact sum rules in eqs. (\ref{cphi}) and 
(\ref{cosdthnu}) correspond to $\sin\theta^e_{23} \neq 0$, 
including the case of a relatively small but non-negligible  
$\sin\theta^e_{23}$. 

 Equation (\ref{s2th12se230phi}) represents an exact sum rule 
connecting the value of the CPV phase $\phi$ with the values of 
the angles $\theta_{13}$ and $\theta_{12}$ for $\theta^e_{23} = 0$.
From eq. (46) we can get approximate sum rules taking into 
account that $\sin\theta_{13} \cong 0.15$:
\begin{align}
\label{s2th12se230appr1}
\sin^2\theta_{12} & =
\sin^2\theta_{12}^\nu
+ \sin 2\theta^\nu_{12}\,\cos\phi\,\sin\theta_{13} 
+ \cos2\theta_{12}^\nu\,\sin^2\theta_{13}
+ O(\sin^4\theta_{13})\,, &\\[0.30cm]
\sin^2\theta_{12} & = \sin^2\theta_{12}^\nu +  
\sin 2\theta^\nu_{12}\,\cos\phi\,\sin\theta_{13}\, + O(\sin^2\theta_{13})\,.
\label{s2th12se230appr2}
\end{align}
%
We have given  the sum rules up to corrections 
of order $O(\sin^4\theta_{13})$ and of order 
$O(\sin^2\theta_{13})$ because both will serve 
our further discussion. By adding and subtracting the 
negligible (within the approximation used)  
term $(\cos\theta^\nu_{12}\,\cos\phi\,\sin\theta_{13})^2$ 
to the r.h.s. of eq. (\ref{s2th12se230appr2}), 
and by using $\sin\theta_{13} \cong \theta_{13}$, 
we get $\sin^2\theta_{12} 
\cong \sin^2(\theta^\nu_{12} + \theta_{13}\cos\phi)$,
which leads to
\begin{equation}
\theta_{12} \cong
 \theta^{\nu}_{12} + \theta_{13}\cos\phi\, + O(\theta^2_{13})\,.
\label{AntKingphi}
\end{equation}
%
 The same result can be obtained 
 by taking the square root of 
 the left-hand and right-hand sides of 
 eq. (\ref{s2th12se230appr2}) treating 
 $\sin2\theta^\nu_{12}\sin\theta_{13}\cos\phi \cong 
 \sin2\theta^\nu_{12}\theta_{13}\cos\phi$ 
 as a small parameter 
 and using the leading order expansion 
 of $\sin^{-1} (p + qx) \cong  
 \sin^{-1}p + qx/\sqrt{1-p^2} + O((qx)^2)$, $qx \ll 1$.

 In what concerns the phase $\delta$,
in the limit of negligible $\theta^e_{23}$ 
we find from eqs. (\ref{cosdthnu}) and (\ref{cosdcosphi}) 
the following exact expressions for 
$\cos\delta$ and the relation between 
$\cos\delta$ and $\cos\phi$:
\begin{align}
\label{cosd23e0} 
\cos\delta & = \; \frac{(1 - 2\sin^2\theta_{13})^{\frac{1}{2}}}
{\sin2\theta_{12}\sin\theta_{13}}\,
\left [\cos2\theta^{\nu}_{12} + 
\left (\sin^2\theta_{12} - \cos^2\theta^{\nu}_{12} \right )\,
\frac{1-3\sin^2\theta_{13}}{1 - 2\sin^2\theta_{13}}\right ]\,, & \\[0.30cm]
\cos\delta & = \; \frac{\sin2\theta^{\nu}_{12}}{\sin2\theta_{12}}\,\cos\phi\,
\frac{1 - 3\sin^2\theta_{13}}{1-\sin^2\theta_{13}}
+ 2\,\frac{\cos2\theta^{\nu}_{12}}{\sin2\theta_{12}}\,
\frac{(1-2\sin^2\theta_{13})^{\frac{1}{2}}}{1-\sin^2\theta_{13}}\,
\sin\theta_{13}\,.
\label{cosdcosphi23e0}
\end{align}
%
Equation (\ref{cosd23e0}) can also be cast in the form of 
eqs. (\ref{s2th12se230phi}), (\ref{s2th12se230appr1}) and 
(\ref{s2th12se230appr2}):
\begin{align}
\label{s2th12se230delta}
\sin^2\theta_{12} & = 
\sin^2\theta_{12}^\nu +  
\frac{(1 - 2\sin^2\theta_{13})^{\frac{1}{2}}}
{1 - 3\sin^2\theta_{13} }\,\sin 2\theta_{12}\, \sin\theta_{13}\,\cos\delta
- \,\frac{\sin^2\theta_{13}}{1-3\sin^2\theta_{13}}\,\cos2\theta_{12}^\nu
 &\\[0.25cm]
\label{s2th12se230deltaappr1}
& = \sin^2\theta_{12}^\nu + 
(1 + 2\sin^2\theta_{13}) \sin2\theta_{12}\cos\delta\,\sin\theta_{13} 
 - \cos2\theta_{12}^\nu \sin^2\theta_{13} + O(\sin^4\theta_{13})\\[0.25cm] 
& = \sin^2\theta_{12}^\nu + 
 \sin 2\theta_{12}\,\cos\delta\,\sin\theta_{13} + O(\sin^2\theta_{13})\,.
\label{s2th12se230deltaappr2}
\end{align}
%
We note that eqs. (\ref{cosd23e0}) - (\ref{s2th12se230delta})
are exact. We have given the approximate sum rules involving 
$\cos\delta$ up to corrections of order 
$O(\sin^4\theta_{13})$ and of order  $O(\sin^2\theta_{13})$  
in eqs. (\ref{s2th12se230deltaappr1}) and 
(\ref{s2th12se230deltaappr2}) because both sum rules 
will be used in the analysis which follows. 

 It is not difficult to show, 
using the same steps which allowed us to get 
eq. (\ref{AntKingphi}) from 
eq. (\ref{s2th12se230appr2}) 
that, to leading order in $\theta_{13}$,
the sum rule in eq. (\ref{s2th12se230deltaappr2}) leads to 
\begin{equation}
\theta_{12} \cong
\theta^{\nu}_{12} + \theta_{13}\cos\delta\, + O(\theta^2_{13})\,.
\label{AntKingdelta}
\end{equation}
%
This implies that, to leading order in 
$\sin\theta_{13}$, the sum rule in 
eq. (\ref{s2th12se230appr2}) is equivalent 
to the sum rule in eq. (\ref{s2th12se230deltaappr2}), 
and thus, to leading order in  $\sin\theta_{13}$, we have 
$\cos\phi \cong \cos\delta$ ~
\footnote{The change  $\sin\theta^e_{12}\rightarrow -\,\sin\theta^e_{12}$
in eq. (9) would lead to the relation $\cos\phi \cong -\,\cos\delta$, 
which appears in a number of articles
(see, e.g., \cite{Antusch:2005kw,Marzocca:2011dh,Antusch:2011qg}).}.
The different expressions in eqs. (\ref{s2th12se230appr2}) 
and (\ref{s2th12se230deltaappr2}) lead to the same 
leading order sum rules (\ref{AntKingphi}) and 
(\ref{AntKingdelta}) as a consequence of the fact 
that the neglected corrections in the two cases differ.

 The approximate sum rules given in 
eqs. (\ref{s2th12se230appr2}), (\ref{AntKingphi}) 
and (\ref{AntKingdelta}) and similar relations, 
were considered or found in specific models,
for different fixed symmetry forms of 
$\tilde{U}_{\nu}$ (BM, TBM, etc.), e.g., in 
\cite{GTani02,FPR04,SPWR04,Hochmuth:2007wq,Marzocca:2011dh,Alta,Antusch:2005kw,
Antusch:2007rk,Antusch:2011qg}~
\footnote{In the BM case, for instance, 
eq. (\ref{s2th12se230appr2}) can be obtained 
i) from eq. (32) in \cite{FPR04}  
by setting the parameters $A = B = 0$,
ii) from eqs. (31)-(32) in the first article 
quoted in \cite{Alta} by setting the parameter 
$s^e_{13}=0$. 
}.
For arbitrary fixed value of $\theta^{\nu}_{12}$ 
the sum rule in eq. (\ref{AntKingphi}) 
was proposed in \cite{Antusch:2005kw},
where the approximate relation 
$\cos\phi \cong \cos\delta$,
which holds to leading order in $\sin\theta_{13}$,
was implicitly used (see further).
 The approximation $|\cos\phi| \cong |\cos\delta|$ 
is employed also, e.g., in refs. 
\cite{Hochmuth:2007wq,Marzocca:2011dh,Antusch:2011qg}. 
It was suggested in  ref. \cite{Antusch:2007rk}
that the sum rule (\ref{AntKingdelta}) 
should be used to obtain the value of $\cos\delta$ 
using the experimentally determined values of 
$\sin^2\theta_{12}$ and $\sin\theta_{13}$, e.g., 
in the case of the TBM form of 
$\tilde{U}_{\nu}$. The same sum rule 
(\ref{AntKingdelta}) is given
also, e.g., in the review articles 
\cite{King:2013eh,King:2014nza}. 

 The derivation of the sum rule 
of interest, given in  ref. \cite{Antusch:2005kw}, 
is based on the following expression for $\sin^2\theta_{12}$:
\begin{eqnarray}
\label{AKs2th120}
\sin^2\theta_{12} \cong 
\left | \sin\theta^{\nu}_{12} + 
 \cos\theta^{\nu}_{12}\sin\theta_{13}\,
e^{i(\delta^{\nu}_{12} - \delta^{e}_{12} + \pi )} \right |^2 \\ [0.25cm] 
\hskip 0.6cm \cong \sin^2\theta^{\nu}_{12} + 
\sin2\theta^\nu_{12}\,\sin\theta_{13}\,
\cos\left ( \delta^\nu_{12} -\delta^e_{12} + \pi \right )\,,
\label{AKs2th12}
\end{eqnarray}
%
where 
\footnote{Expression (\ref{AKs2th12}) follows 
from eqs. (15c) and (18) in ref. \cite{Antusch:2005kw} after, 
following ref.  \cite{Antusch:2005kw},
one neglects the term   $\propto \theta^e_{13}$ in eq. (15c) and
uses $c^{\nu}_{23} = s^{\nu}_{23}$.}
$\delta^\nu_{12}$ and  $\delta^e_{12}$ are two
of the phases introduced in  \cite{Antusch:2005kw}.
The expression for $\sin^2\theta_{12}$
in eq. (\ref{AKs2th12}) is obtained  in \cite{Antusch:2005kw}
by keeping the leading order corrections in
$\sin\theta_{13}$ and  $\sin\theta^e_{23}\neq 0$, 
$\sin\theta^e_{23}\ll 1$, 
and neglecting terms of order of, or smaller than,  
$\sin^2\theta_{13}$, $\sin^2\theta^e_{23}$ and
$\sin\theta_{13}\sin\theta^e_{23}$. 
The presence of  $\sin2\theta^\nu_{12}$ 
(rather than $\sin2\theta_{12}$) in eq. (\ref{AKs2th12}) 
suggests a similarity between this equation and 
eq. (\ref{s2th12se230appr2}) in which the phase 
$\phi$ is present. 
It can be shown that the following exact relation holds 
between the phase $\psi$, defined in eq. (\ref{PsieQ0}), 
and the phases  $\delta^\nu_{12}$ and  $\delta^e_{12}$, 
introduced
\footnote{The relations between the phases 
$(\delta^\nu_{12} -\delta^e_{12})$ and $\psi$ or $\phi$ 
we are going to derive 
are valid, obviously, modulo 2$\pi$.} 
in  \cite{Antusch:2005kw}:
\begin{equation} 
\psi = -\, \left ( \delta^\nu_{12} -\delta^e_{12} + \pi \right )\,.
\label{psiAKdeltas}
\end{equation}
%
Further, it follows from eq. (\ref{gammaphi}) that the phases 
$\psi$ and $\phi$ are related in the following way:
\begin{equation}
\psi = -\, (\phi - \tilde\phi)\,,
\label{psiphi}
\end{equation}
%
where 
\begin{equation}
\sin\tilde\phi = \frac{\sin\theta^e_{23}\,\sin(\psi - \omega)} 
{\sqrt{1 + \sin2\theta^e_{23}\,\cos(\psi - \omega)}}\,. 
\label{sintildephi}
\end{equation}
%
We note that in the absence of 
1-3 rotations in $\tilde{U}_e$ and $\tilde{U}_{\nu}$,
the relations (\ref{psiAKdeltas}) - (\ref{sintildephi}) 
are exact. It follows from 
eqs. (\ref{psiAKdeltas}) -  (\ref{sintildephi}) that for the 
phase $(\delta^\nu_{12} - \delta^e_{12} + \pi)$  
in eq. (\ref{AKs2th12}) we get:
\begin{equation}
\delta^\nu_{12} - \delta^e_{12} + \pi = \phi - \tilde\phi\,.  
\label{AKdeltasphi}
\end{equation}
%
This implies that, in the approximation employed 
in  ref. \cite{Antusch:2005kw} in which terms of order 
$\sin\theta_{13}\sin\theta^e_{23}$ are neglected,
the contribution of 
$\tilde\phi$ in eq. (\ref{AKs2th12}) should also 
be neglected and we get:
\begin{eqnarray}
\sin^2\theta_{12} \cong
\sin^2\theta^{\nu}_{12}  + 
\sin2\theta^\nu_{12}\,\sin\theta_{13}\,\cos\phi\,,
\label{AKs2th122}
\end{eqnarray}
%
which coincides with eq. (\ref{s2th12se230appr2}) in which 
the phase $\phi$, rather than the phase $\delta$, 
is present. 

 In the case of $\theta^e_{23} =0$ we get from 
eqs. (\ref{sintildephi}) and  (\ref{AKdeltasphi}) 
the exact relation:
\begin{equation}
\delta^\nu_{12} - \delta^e_{12} + \pi = \phi\,.  
\label{AKdeltasphiexact}
\end{equation}
%
We find the same relation comparing the expressions for the 
rephasing invariant $J_{cP}$, eq. (\ref{JCP}), 
in the standard parametrisation of the PMNS matrix and 
in the parametrisation employed 
in ref. \cite{Antusch:2005kw}. This allows us to obtain a 
relation between the phase $\delta$ and the 
phase $(\delta^\nu_{12} - \delta^e_{12})$, which in turn, 
via eq.  (\ref{sindsinphi}), leads to a relation between 
$(\delta^\nu_{12} - \delta^e_{12})$ and $\phi$.
Indeed, taking into account that in the case of 
$\theta^e_{23} =0$, $\sin^2\theta_{23}$ 
is given in eq. (\ref{s2th13se230}), and that in the 
parametrisation used in  \cite{Antusch:2005kw} one has 
$\theta^{\nu}_{23} = \pi/4$, $\sin\theta_{13}= \sin\theta^e_{12}/\sqrt{2}$, 
we get equating the two expressions of interest 
for the $J_{cP}$ factor:
\begin{equation}
\sin\delta = -\, \frac{\sin2\theta^{\nu}_{12}}{\sin2\theta_{12}}\,
\sin(\delta^\nu_{12} - \delta^e_{12} + \pi)\,.
\label{sindsinAKds}
\end{equation}
%
This result is exact. Comparing the above equation with 
eq. (\ref{sindsinphi}) we can conclude that 
\begin{equation}
\sin(\delta^\nu_{12} - \delta^e_{12} + \pi) =  \sin\phi\,,
\label{sinphisinAKds}
\end{equation}
%
which leads to eq. (\ref{AKdeltasphiexact}) 

As we have already noted,  in the derivation of the sum rule 
under discussion 
proposed in \cite{Antusch:2005kw},  terms of order 
$\sin^2\theta_{13}$, $\sin^2\theta^e_{23}$ and
$\sin\theta_{13}\sin\theta^e_{23}$ and higher order 
corrections are neglected. 
In the next subsection we will consider 
the corrections due to $\sin\theta^e_{23}\neq 0$.
Here we would like to note that for the  
TBM, GRA, GRB and HG 
forms of the matrix $\tilde{U}_{\nu}$ of interest, we have 
$|\sin^2\theta_{12} - \sin^2\theta^{\nu}_{12}|\sim \sin^2\theta_{13}$.
Indeed, for the best fit value of 
$\sin^2\theta_{12} = 0.308$, this difference in the TBM, 
GRA, GRB and HG cases  reads, respectively: 
0.032;~0.025;~0.037;~0.058. 
Therefore in all four cases under discussion we have 
$\sin^2\theta_{12} = \sin^2\theta^{\nu}_{12} + a \sin^2\theta_{13}$, 
with $|a| \cong (1.1-2.5)$. The last relation implies:
\begin{equation}
\theta_{12} = \theta^{\nu}_{12} + 
 \frac{a\,\theta^2_{13}}{\sin2\theta^{\nu}_{12}} + 
O(a^2\theta^4_{13})\,,~~~1.1\ltap |a| \ltap 2.5\,,
\label{th12th12nu}
\end{equation}
%
where $\sin2\theta^{\nu}_{12} \cong 0.94$, 0.89, 0.95 and 0.87 
for the TBM, GRA, GRB and HG forms of $\tilde{U}_{\nu}$, 
respectively. Thus, if one should be consistent, 
working in the leading order approximation
in $\sin\theta_{13}$, i.e., neglecting 
terms $\sim \sin^n\theta_{13}$ for $n\geq 2$, 
for the TBM, GRA, GRB and HG forms of 
$\tilde{U}_{\nu}$, one should also neglect 
the difference between $\sin^2\theta_{12}$ and 
$\sin^2\theta^{\nu}_{12}$ in  eqs. (\ref{s2th12se230appr2}) and 
(\ref{s2th12se230deltaappr2}), or equivalently, 
the difference between $\theta_{12}$ and $\theta^{\nu}_{12}$ 
in  eqs. (\ref{AntKingphi}) and (\ref{AntKingdelta}).
In this case we get $\cos\phi = \cos\delta$, 
but also $\cos\phi = 0$ and   $\cos\delta =0$, 
for the indicated symmetry forms of $\tilde{U}_{\nu}$.
If the sum rules are derived in  the TBM, GRA, GRB and HG cases
taking into account the difference 
$|\theta_{12} - \theta^{\nu}_{12}|\sim \theta^2_{13}\neq 0$, 
(or $|\sin^2\theta_{12} - 
\sin^2\theta^{\nu}_{12}| \sim \sin^2\theta_{13} \neq 0$), 
a consistent application of the approximations 
used requires in  
these cases to take also terms of order 
$\sin^2\theta_{13}$ into account, i.e., 
to use the sum rules given in eqs. (\ref{s2th12se230appr1}) and 
(\ref{s2th12se230deltaappr1}), rather than the sum rules 
(\ref{s2th12se230appr2}) and (\ref{s2th12se230deltaappr2})  
(or (\ref{AntKingphi}) and (\ref{AntKingdelta})).
We will use quotation marks in the term 
``leading order sum rules'' 
to denote the inconsistency of the approximations 
used to derive the sum rules in eqs. (\ref{s2th12se230appr2}) 
and (\ref{s2th12se230deltaappr2}), and, correspondingly, in eqs.  
(\ref{AntKingphi}) and (\ref{AntKingdelta}), 
in the TBM, GRA, GRB and HG cases.
We will return to the problem of correct 
implementation of  the approximations employed 
to derive the sum rules 
in eqs. (\ref{AntKingphi}) and (\ref{AntKingdelta})
in the next subsection, where we will analyse in detail 
the corrections due to $\sin\theta^e_{23}\neq 0$.

 The above considerations do not apply to 
the case of BM (LC) form of the matrix $\tilde{U}_{\nu}$ 
since in this case we have 
$|\sin^2\theta_{12} - \sin^2\theta^{\nu}_{12}| 
\sim \sin\theta_{13}$. 
Thus, for the  BM (LC) form of  $\tilde{U}_{\nu}$,
the leading order approximation 
in $\sin\theta_{13}$ is consistent with 
taking into account the difference between 
$\theta_{12}$ and  $\theta^{\nu}_{12}$
in the sum rules given in 
eqs. (\ref{AntKingphi}) and (\ref{AntKingdelta}), 
and in eqs. (\ref{s2th12se230appr2}) 
and (\ref{s2th12se230deltaappr2}).  

 We will show next that the sum rules 
in eqs. (\ref{s2th12se230appr2}) and  
(\ref{s2th12se230deltaappr2}), and the equivalent 
``leading order sum rules'' in eqs. (\ref{AntKingphi}) 
and (\ref{AntKingdelta}), give imprecise, 
and in some cases - largely incorrect,
results for both $\cos\phi$ and $\cos\delta$ 
in the cases of TBM, GRA, GRB and HG 
forms of $\tilde{U}_{\nu}$.

 Indeed, using the ``leading order sum rules'' 
in eqs. (\ref{s2th12se230appr2}) and 
(\ref{s2th12se230deltaappr2}), we get 
for the best fit values of 
$\sin^2\theta_{12} = 0.308$ and 
$\sin^2\theta_{13} = 0.0234$ 
in the TBM, GRA, GRB and HG cases 
\footnote{Practically the same results are obtained employing the 
equivalent ``leading order sum rules'' in eqs.  
(\ref{AntKingphi}) and (\ref{AntKingdelta}).}: 
\begin{align}
\label{TBMappr2}
{\rm TBM,~eqs.~(\ref{s2th12se230deltaappr2})~and~ 
(\ref{s2th12se230appr2}):}
~~\cos\delta & = -\,0.179\,;~~\cos\phi \cong -\,0.176\,;
& \\[0.25cm]
\label{GRAappr2}
{\rm GRA,~eqs.~(\ref{s2th12se230deltaappr2})~and~ 
(\ref{s2th12se230appr2}):}
~~\cos\delta & \cong 0.227\,;~~\cos\phi \cong 0.234\,;
& \\[0.25cm]
\label{GRBappr2}
{\rm GRB,~eqs.~(\ref{s2th12se230deltaappr2})~and~ 
(\ref{s2th12se230appr2}):}
~~\cos\delta & \cong -\,0.262\,;
            ~~~~\cos\phi \cong -\,0.254\,;
& \\[0.25cm]
\label{HGappr2}
{\rm HG,~eqs.~(\ref{s2th12se230deltaappr2})~and~ 
(\ref{s2th12se230appr2}):}
~~\cos\delta & \cong 0.411\,;~~\cos\phi \cong 0.438\,.
\end{align}
%
 Clearly, in all these cases we have 
$\cos\delta \cong \cos\phi$.
The slight differences in the values of 
$\cos\delta$ and $\cos\phi$ are caused by the 
differences between the factors 
$\sin2\theta^{\nu}_{12}$ and $\sin2\theta_{12}$
in eqs. (\ref{s2th12se230appr2}) and 
(\ref{s2th12se230deltaappr2}).
In the approximation in which 
eqs. (\ref{AntKingphi}) and (\ref{AntKingdelta}) 
are derived, these differences
should be neglected and we would have 
$\sin2\theta^{\nu}_{12} = \sin2\theta_{12}$.
But in this case, as we have already have noticed, 
we would have also $\theta^{\nu}_{12} = \theta_{12}$, 
and thus $\cos\delta = \cos\phi = 0$.

  Using the exact sum rules 
for $\cos\phi$ and $\cos\delta$, 
given in eqs. (\ref{cosd23e0}) and (\ref{s2th12se230phi}), 
we find:
\begin{align}
\label{TBMexact}
{\rm TBM~exact~:}~~~~\cos\delta & = -\,0.114\,;~~~~\cos\phi \cong -\,0.230\,;
& \\[0.25cm]
\label{GRAexact}
{\rm GRA~exact~:}~~~~\cos\delta & \cong 0.289\,;~~~~\cos\phi \cong 0.153\,;
& \\[0.25cm]
\label{GRBexact}
{\rm GRB~exact~:}~~~~\cos\delta & \cong -\,0.200\,;
            ~~~~\cos\phi \cong -\,0.307\,;
& \\[0.25cm]
\label{HGexact}
{\rm HG~exact~:}~~~~\cos\delta & \cong 0.476\,;~~~~\cos\phi \cong 0.347\,.
\end{align}
%
As we see comparing eqs. (\ref{TBMappr2}) - (\ref{HGappr2}) 
with eqs. (\ref{TBMexact}) - (\ref{HGexact}), 
the values of $\cos\delta$, 
obtained using the exact 
sum rule (\ref{cosd23e0}) in the 
TBM,  GRA, GRB and HG cases 
differ from those calculated 
using the ``leading order sum rule'' 
(\ref{s2th12se230deltaappr2}), 
by the factors 
1.57, 0.78, 1.31 and 0.86, 
respectively. In the case of 
$\cos\phi$, the corresponding factors are 
0.76, 1.53, 0.83 and 1.26. 
The higher order corrections 
have opposite effect on the leading order 
results for $|\cos\delta|$ and $|\cos\phi|$: 
if the exact sum rule value of 
$|\cos\delta|$ is smaller (larger) 
than the ``leading order sum rule'' value, 
as in the TBM and GRB (GRA and HG) cases, 
the corresponding exact sum rule value of 
$|\cos\phi|$ is larger (smaller) 
than the ``leading order sum rule'' value.
We see also from eqs. (\ref{TBMexact}) - (\ref{HGexact})
that the values of $\cos\delta$ and $\cos\phi$, 
derived from the exact sum rules 
in  the cases of  TBM, GRA, GRB and HG 
forms of the matrix $\tilde{U}_{\nu}$
indeed differ approximately by factors
(1.5 - 2.0).  As we have seen, for finite
values of  $\theta^e_{23}$, for which 
we have $\sin^2\theta_{23} \cong (0.43 - 0.44)$,
$\cos\phi$ and $\cos\delta$ in all cases we are 
considering with the exception of the BM (LC) one, differ 
approximately by the same factor of (1.5 - 2.0).

 The origin of these significant 
differences between the results derived 
using the exact and the ``leading order
sum rules'' for $\cos\delta$ 
and $\cos\phi$ for the TBM, GRA, GRB and HG 
forms of the matrix $\tilde{U}_{\nu}$ 
can be traced to the importance of the
next-to-leading order corrections 
$\propto \sin^2\theta_{13}$ 
in the ``leading order sum rules''  for 
\footnote{Note that since in the sum rules of interest 
$\cos\delta$ and $\cos\phi$ are always multiplied 
by $\sin\theta_{13}$, the corrections $\sim \sin^2\theta_{13}$ 
in the sum rules lead effectively to corrections 
$\sim \sin\theta_{13} \cong 0.16$ in the values of 
$\cos\delta$ and $\cos\phi$.
}
$\cos\delta$ and $\cos\phi$. 
For arbitrary fixed $\theta^{\nu}_{12}$
these corrections are given in 
eqs. (\ref{s2th12se230appr1}) and 
(\ref{s2th12se230deltaappr1}).
In the specific cases of TBM GRA, GRB and HG 
forms of $\tilde{U}_{\nu}$, up to corrections 
$O(\sin^4\theta_{13})$ the sum rules for 
$\cos\delta$ read:
\begin{align}
\label{TBMsumrdelta}
{\rm TBM:}~\sin^2\theta_{12} & \cong
\frac{1}{3}\left (1 - \sin^2\theta_{13} \right) + 
 (1+2\sin^2\theta_{13})\sin2\theta_{12}\sin\theta_{13}\,\cos\delta\,,
 \\[0.25cm]
\label{GRAsumrdelta}
{\rm GRA:}~\sin^2\theta_{12}& \cong
0.276(1 + 2\sin^2\theta_{13}) - \sin^2\theta_{13} + 
(1+2\sin^2\theta_{13})\,\sin2\theta_{12}\sin\theta_{13}\,\cos\delta\,,
\end{align}
\begin{align}
\label{GRBsumrdelta}
{\rm GRB:}~\sin^2\theta_{12}& \cong
0.345(1+2\sin^2\theta_{13}) - \sin^2\theta_{13} + 
(1+2\sin^2\theta_{13})\sin2\theta_{12}\sin\theta_{13}\cos\delta\,,
\\[0.25cm]
\label{HGsumrdelta}
{\rm HG:}~\sin^2\theta_{12} & \cong
\frac{1}{4}\left (1 - 2\,\sin^2\theta_{13} \right) + 
(1+2\sin^2\theta_{13})\,\sin2\theta_{12}\sin\theta_{13}\,\cos\delta\,,
\end{align}
%
where we have used $\sin^2\theta^\nu_{12} = (2 + r)^{-1} \cong 0.276$, 
$\sin^2\theta^{\nu}_{12} = (3-r)/4 \cong 0.345$ 
and  $\sin^2\theta^{\nu}_{12} =1/4$ in the GRA, GRB and HG cases, 
respectively (we recall that $r = (1 + \sqrt{5})/2$ 
is the golden ratio). 
Similarly, for the sum rules involving the  phase $\phi$ 
we find:
\begin{align}
\label{TBMsumrphi}
{\rm TBM:}~\sin^2\theta_{12} & =  
\frac{1}{3}\left (1 + \sin^2\theta_{13} \right) + 
\frac{2\sqrt{2}}{3}\sin\theta_{13}\,\cos\phi\,
 + O(\sin^4\theta_{13})\,, & \\[0.25cm]
\label{GRAsumrphi}
{\rm GRA:}~\sin^2\theta_{12}& \cong 
0.276\cos2\theta_{13} + \sin^2\theta_{13} + 
0.894\,\sin\theta_{13}\, \cos\phi + O(\sin^4\theta_{13})\,,
\end{align}
%
\begin{align}
\label{GRBsumrphi}
{\rm GRB:}~\sin^2\theta_{12} & =
0.345\cos2\theta_{13} + \sin^2\theta_{13} + 
0.951\,\sin\theta_{13}\, \cos\phi + O(\sin^4\theta_{13})\,, & \\[0.25cm]
\label{HGsumrphi}
{\rm HG:}~\sin^2\theta_{12} & = 
\frac{1}{4}\left (1 + 2\,\sin^2\theta_{13} \right) + 
\frac{\sqrt{3}}{2}\sin\theta_{13}\,\cos\phi\,
 + O(\sin^4\theta_{13})\,.
\end{align}
%
As can be easily checked, 
the approximate sum rules 
given in eqs. (\ref{TBMsumrdelta}) - (\ref{HGsumrphi}), 
lead to results for $\cos\delta$ and  $\cos\phi$, 
which practically coincide with those quoted in 
eqs. (\ref{TBMexact}) -  (\ref{HGexact})
and obtained using the exact sum rules given in 
eqs. (\ref{cosd23e0}) and (\ref{s2th12se230phi}).
It follows from eqs. (\ref{s2th12se230appr1}) and 
(\ref{s2th12se230deltaappr1}) that the important 
corrections $\propto \sin^2\theta_{13}$ 
to the ``leading order sum rules'' eqs. (\ref{s2th12se230appr2}) 
and (\ref{s2th12se230deltaappr2}),
are given respectively by $(+\,\cos2\theta^{\nu}_{12} \sin^2\theta_{13})$
and by  $(-\,\cos2\theta^{\nu}_{12} \sin^2\theta_{13})$, 
i.e., they coincide in absolute value but have opposite 
signs. This explains the effect of these corrections 
on the values of $|\cos\delta|$ and $|\cos\phi|$ 
derived from the ``leading order sum rules'' 
(\ref{s2th12se230appr2}) and (\ref{s2th12se230deltaappr2}): 
given the value of  $|\cos2\theta^{\nu}_{12} \sin^2\theta_{13}|$, 
the corrections make maximal the difference between 
$|\cos\delta|$ and $|\cos\phi|$.
The fact that the correction $\propto \sin^2\theta_{13}$ of interest 
is given by the term  $\pm\,\cos2\theta^{\nu}_{12} \sin^2\theta_{13}$ 
explains also why the results for 
$\cos\delta$ and $\cos\phi$ obtained using 
the exact sum rules  (\ref{cosd23e0}) and (\ref{s2th12se230phi})
and leading order sum rules 
(\ref{s2th12se230appr2}) and (\ref{s2th12se230deltaappr2}) 
do not differ significantly for BM (LC) form 
of the matrix $\tilde{U}_{\nu}$: in the BM (LC) case 
these correction is zero since
$\cos2\theta^{\nu}_{12} = 0$.
Thus, the corrections to the leading order sum rule 
are $O(\sin^3\theta)$ and  $O(\sin^4\theta)$
and have minor effect on the determination of 
$\cos\delta$ and $\cos\phi$ in the BM (LC) case.

 We would like to emphasise once again that 
the corrections $\propto \sin^2\theta_{13}$ to
the ``leading order sum rules''for $\cos\delta$ and $\cos\phi$ 
 (\ref{AntKingdelta}) and (\ref{AntKingphi}), as well as, 
(\ref{s2th12se230deltaappr2}) and (\ref{s2th12se230appr2}), 
are significant and have to be taken into account 
when the difference 
$|\sin^2\theta_{12} - \sin^2\theta^{\nu}_{12}|\sim \sin^2\theta_{13}$, 
and thus is of the order of the correction. 
For the current best fit value of $\sin^2\theta_{12} = 0.308$ 
this is the case of the TBM, GRA, GRB and HG forms 
of the matrix $\tilde{U}_{\nu}$ considered in the present article. 
%
\subsection{The Corrections Generated by
Non-negligible $\sin\theta^e_{23}\ll 1$}
%
%
 The sum rule (\ref{AKs2th120}), which leads to the 
 ``leading order sum rules'' 
(\ref{AntKingphi}) and (\ref{AntKingdelta})
of interest, was derived in \cite{Antusch:2005kw} 
assuming that $\theta^e_{12} \neq 0$, 
$\theta^e_{23}\neq 0$ and $|\sin\theta^e_{23}| \ll 1$, 
and keeping terms $\sim\sin\theta_{13}$ and $\sim \sin\theta^e_{23}$ 
in the relation between $\sin\theta_{12}$, 
$\sin\theta^{\nu}_{12}$ and $\cos\delta$.
The corrections of the order of, or smaller than, 
$\sin^2\theta_{13}$, $\sin^2\theta^e_{23}$ and 
$\sin\theta_{13}\sin\theta^e_{23}$ were 
neglected. The exact sum rules for 
$\cos\phi$ and $\cos\delta$ given 
in  eqs. (\ref{cphi}) and (\ref{cosdthnu}), 
were derived for any $\theta^e_{12} \neq 0$, 
$\sin\theta_{13}$ and $\sin\theta^e_{23}$.  
Thus, the sum rule (\ref{AntKingdelta}) is 
an approximate version of the exact sum rule 
(\ref{cosdthnu}): eq. (\ref{AntKingdelta}) 
can be obtained from eq. (\ref{cosdthnu}) 
in the leading order approximation 
by treating not only $\sin\theta_{13}$, 
but also $\sin\theta^e_{23}$ as a 
small parameter.  In this subsection, 
from the exact sum rules (\ref{cphi}) and (\ref{cosdthnu}), 
we will derive the corrections due to 
both $\sin\theta_{13}$ and  $\sin\theta^e_{23}\neq 0$ in the 
``leading order sum rules'' in   
eqs. (\ref{s2th12se230appr2}) and (\ref{s2th12se230deltaappr2}), 
and in eqs. (\ref{AntKingphi}) and (\ref{AntKingdelta}).

 It follows from eq. (\ref{th23hat}) that 
\begin{equation}
\sin2\theta^e_{23}\cos(\omega - \psi) \equiv X = 
1 - 2\sin^2\hat\theta_{23} \cong 0.124\,,
\label{X}
\end{equation}
%
The relation between 
$\sin^2\theta_{23}$ and $\sin^2\hat\theta_{23}$ is 
given in eq. (\ref{s2th23}). The numerical value 
quoted in eq. (\ref{X}) is for 
$\sin^2\hat\theta_{23} \cong 0.438$, which 
corresponds to 
$\sin^2\theta_{23} =0.425$ and $\sin^2\theta_{13} =0.0234$. 

 Equation (\ref{X}) implies  that 
$|\sin2\theta^e_{23}|\gtap 0.124$. 
Following the analysis performed in 
\cite{Antusch:2005kw}, we will assume that 
$0 < \sin\theta^e_{23} \ll 1$, and thus $X \ll 1$.
From the exact sum rules for 
$\cos\phi$ and $\cos\delta$ given 
in  eqs. (\ref{cphi}) and (\ref{cosdthnu}), 
we will derive approximate sum rules 
for the two CPV phases, in which, in contrast 
to the approximation employed in 
ref. \cite{Antusch:2005kw} leading to 
eq. (\ref{AntKingdelta}),  
the next-to-leading order corrections   
$\sim\sin^2\theta_{13}$, $\sim\sin^2\theta^e_{23}$ and 
$\sim\sin\theta_{13}\sin\theta^e_{23}$ are included.
This means that, in addition to keeping terms 
$\sim \sin\theta_{13}$ and $\sim \sin^2\theta_{13}$
in the sum rules, we will keep also 
terms $\sim X$,  $\sim X^2$ and 
$\sim X\sin\theta_{13}$. 
It is not difficult to show that 
in this next-to-leading order 
approximation we get from 
eqs. (\ref{cosdthnu}) and (\ref{cphi}):
\begin{eqnarray}
\nonumber
\sin^2\theta_{12} = \sin^2\theta_{12}^\nu + 
(1 + X)\sin 2\theta_{12}\, \sin\theta_{13}\,\cos\delta
- \,\cos2\theta_{12}^\nu\, \sin^2\theta_{13} \\[0.25cm]
\hskip 3cm  
+ \, O(X\sin^2\theta_{13},X^2\sin\theta_{13},\sin^3\theta_{13},X^3)\,,
 \label{s2th12deltaappr3X}
\\[0.25cm]
\nonumber
\sin^2\theta_{12} =
\sin^2\theta_{12}^\nu
+ (1+X)\,\sin 2\theta^\nu_{12}\,\cos\phi\,\sin\theta_{13} 
+ \cos2\theta_{12}^\nu\,\sin^2\theta_{13}\\[0.25cm]
\hskip 3cm 
+ \, O(X\sin^2\theta_{13},X^2\sin\theta_{13},\sin^3\theta_{13},X^3)\,.
\label{s2th12phiappr3X}
\end{eqnarray}
%
Comparing eqs. (\ref{s2th12deltaappr3X}) and (\ref{s2th12phiappr3X})
respectively 
with eqs. (\ref{s2th12se230deltaappr1}) and (\ref{s2th12se230appr1}),  
we see that the next-to-leading order correction 
due to $X \sim \sin\theta^e_{23}\neq 0$ 
amounts formally to multiplying  
the terms $\propto \sin\theta_{13}\cos\delta$ 
and $\propto \sin\theta_{13} \cos\phi$ by 
the factor $(1 + X)$. The
``leading order sum rules'' in   
eqs. (\ref{s2th12se230appr2}) and (\ref{s2th12se230deltaappr2}), 
and in eqs. (\ref{AntKingphi}) and (\ref{AntKingdelta}), 
do not depend on $\sin\theta^e_{23}$ because 
in the sum rules (\ref{s2th12deltaappr3X}) and 
(\ref{s2th12phiappr3X})
there are no terms of the order 
of  $\sin\theta^e_{23}$: the small 
parameter $\sin\theta^e_{23}$ appears only in 
the next-to-leading order correction 
$\sim\sin\theta^e_{23}\sin\theta_{13}$.

It follows from eqs. (\ref{X}) and 
(\ref{s2th23}) that we have:
$ 1 + X = 2\cos^2\hat\theta_{23} = 
2\cos^2\theta_{23}(1 - \sin^2\theta_{13})$.
Thus, in the approximation of interest the sum rules 
for $\cos\delta$ and $\cos\phi$ take the form:
\begin{eqnarray}
\nonumber
\sin^2\theta_{12} = \sin^2\theta_{12}^\nu + 
2\cos^2\theta_{23}\,\sin 2\theta_{12}\, \sin\theta_{13}\,\cos\delta
- \,\cos2\theta_{12}^\nu\, \sin^2\theta_{13} \\[0.25cm]
\hskip 3cm 
+ \, O(X\sin^2\theta_{13},X^2\sin\theta_{13},\sin^3\theta_{13},X^3)\,,
 \label{s2th12deltaappr3}
\\[0.25cm]
\nonumber
\sin^2\theta_{12} =
\sin^2\theta_{12}^\nu
+ 2\cos^2\theta_{23}\,\sin 2\theta^\nu_{12}\,\cos\phi\,\sin\theta_{13} 
+ \cos2\theta_{12}^\nu\,\sin^2\theta_{13}  \\[0.25cm]
\hskip 3cm  
+ \, O(X\sin^2\theta_{13},X^2\sin\theta_{13},\sin^3\theta_{13},X^3)\,.
\label{s2th12phiappr3}
\end{eqnarray}
%
 For $\theta^e_{23} = 0$, we have  $2\cos^2\theta_{23} = 
(1 - \sin^2\theta_{13})^{-1}$, 
and, within the approximation employed, 
eqs. (\ref{s2th12deltaappr3}) and (\ref{s2th12phiappr3})
reduce to eqs. (\ref{s2th12se230deltaappr1}) and (\ref{s2th12se230appr1}).
In the case of non-negligible $\theta^e_{23}$, however, 
$\sin^2\theta_{23}$ can deviate sizably from 0.5.
In this case, as it follows from eqs. (\ref{cosdthnu}),
(\ref{cphi}), (\ref{s2th12deltaappr3}) 
and (\ref{s2th12phiappr3}), 
the exact and the approximate 
(next-to-leading order)  sum rules for  
$\cos\delta$ and $\cos\phi$ 
depend not only on $\theta_{12}$ and $\theta_{13}$, 
but also on $\theta_{23}$. 
If, for instance, $\cos^2\theta_{23} = 0.6~(0.4)$, 
the effect of the factor  $2\cos^2\theta_{23}$, e.g.,
in the approximate sum rules (\ref{s2th12deltaappr3}) 
and (\ref{s2th12phiappr3}) is to 
decrease (increase) the values of 
$\cos\delta$ and $\cos\phi$, 
evaluated without taking into account the 
correction due to $\theta^e_{23}\neq 0$,
by a factor of 1.2 (1.25).  
This dependence, as well as the  variation of the 
predictions for $\cos\delta$ and  $\cos\phi$ with the 
variation of the values of 
$\sin^2\theta_{12}$, $\sin^2\theta_{23}$ 
and $\sin^2\theta_{13}$ in their experimentally 
allowed ranges, will be investigated elsewhere  
\cite{Girardi:2014xyz}.

\vspace{-0.2cm}
%
\section{The Majorana Phases}
\label{Sec:MajPhs}
%

 We will analyse next the possibility to obtain predictions for 
the values of Majorana phases $\alpha_{21}$ and $\alpha_{31}$
in the PMNS matrix using the approach described above, 
We will show in what follows that in many cases 
of interest it is possible to determine the phases 
 $\alpha_{21}$ and $\alpha_{31}$ if the values of 
the phase $\phi$, or $\delta$, and 
of the phases $\xi_{21}$ and $\xi_{31}$ 
in the diagonal matrix $Q_0$ in eq. (\ref{Ugeneral}) 
are known. The matrix  $\tilde{U}_\nu Q_0$, 
as we have already briefly discussed, 
originates from the diagonalisation of the 
flavour neutrino Majorana mass term. 
In many theories and models of neutrino mixing
the values of the phases  $\xi_{21}$ and $\xi_{31}$ 
are fixed by the form of flavour neutrino 
Majorana mass term, which is dictated by the 
chosen discrete (or continuous) flavour symmetry
(see, e.g., \cite{Chen:2009gf,Girardi:2013sza}), 
or on phenomenological grounds (see, e.g.,\cite{Duarah:2012bm}).
Typical values of the phases  
$\xi_{21}$ and $\xi_{31}$ are 0, $\pi/2$ and $\pi$. 
In the model with $T^\prime$ flavour symmetry in the lepton sector 
constructed in \cite{Girardi:2013sza}, for instance, 
$\xi_{21}$ and $\xi_{31}$ can take two sets of values:
$(\xi_{21},\xi_{31}) = (0,0)$ and $(0,\pi)$. 

 In what follows we will assume that the  phases 
$\xi_{21}$ and $\xi_{31}$ are known.
Under this condition the Majorana phases 
$\alpha_{21}$ and $\alpha_{31}$ can be determined, 
as we will discuss in greater detail below, 
i) if the angles  $\theta^e_{12}$ and 
$\theta^e_{23}$, or the angle  $\theta^e_{12}$ 
and the phase $(\psi - \omega)$, are known, or
ii) if the angle  $\theta^e_{12}$ is known and the 
phase $\psi$ or $\omega$ takes one of the specific values 
0, $\pi/2$, $\pi$ and $3\pi/2$.
In processes like the $\betabeta$-decay,  
which are characteristic of the Majorana nature 
of the light massive neutrinos $\nu_j$,
the phase $\alpha_{31}$ can play 
under certain conditions a subdominant role 
(see further), while the 
rate of the processes depends strongly 
on the phase  $\alpha_{21}$. 
As we will see, the phase  $\alpha_{21}$
can be determined (given the phase 
$\xi_{21}$) knowing only 
the values of the phase $\phi$ (or $\delta$) 
and of the angle  $\theta^e_{12}$.

 The PMNS matrix we obtain from 
eq. (\ref{UPMNSthhat1}) in the scheme 
we consider has the form:
\begin{equation} 
\begin{array}{c}
\label{Uthenuhat}
U_{\rm PMNS} = \left(\begin{array}{ccc} 
c^e_{12} c^\nu_{12} -  s^e_{12}\hat{c}_{23} s^\nu_{12} e^{\ci \phi}& 
 c^e_{12} s^\nu_{12} + s^e_{12}\hat{c}_{23} c^\nu_{12} e^{\ci \phi}& 
s^e_{12} \hat{s}_{23} e^{\ci \phi}  \\[0.2cm] 
- s^e_{12} c^\nu_{12} -  c^e_{12}\hat{c}_{23} s^\nu_{12} e^{\ci \phi}& 
- s^e_{12} s^\nu_{12} +  c^e_{12}\hat{c}_{23} c^\nu_{12} e^{\ci \phi} & 
c^e_{12} \hat{s}_{23} e^{\ci \phi} \\[0.2cm] 
\hat{s}_{23}s^\nu_{12} &  - \hat{s}_{23} c^\nu_{12} 
 & \hat{c}_{23} 
\\ 
\end{array}    
\right)\, Q_1\,Q_0\,,
\end{array} 
\end{equation}
%
where we have used the standard notations 
$c^e_{12} \equiv \cos\theta^e_{12}$, 
$c^\nu_{12} \equiv \cos\theta^\nu_{12}$,
$\hat{c}_{23} \equiv \cos\hat\theta_{23}$, etc.
Obviously, the matrix (\ref{Uthenuhat}) does not have the form of 
the standard parametrisation of the PMNS matrix. 
As we will show below, bringing the matrix (\ref{Uthenuhat}) 
to the standard parametrisation form leads 
to contributions to the Majorana phases $\alpha_{21}$ 
and $\alpha_{31}$, which are associated with the phase 
$\phi$. Thus, the phase $\phi$ not only 
generates the Dirac phase $\delta$, but also contributes 
to the values of the Majorana phases $\alpha_{21}$ 
and $\alpha_{31}$.

 The first thing to notice is that 
using eqs. (\ref{s2th13}) - (\ref{s2th12}) it can be shown  
that the absolute values of the elements of the matrix 
given in eq. (\ref{Uthenuhat}) coincide with the 
absolute values of the elements of the PMNS matrix 
in the standard parametrisation, defined 
in eqs. (\ref{eq:VQ}) - (\ref{eq:Vpara}):    
$|c^e_{12} c^\nu_{12} -  s^e_{12}\hat{c}_{23} s^\nu_{12} e^{i \phi}| = 
c_{12}c_{13} = |U_{e1}|$,
$|c^e_{12} s^\nu_{12} + s^e_{12}\hat{c}_{23} c^\nu_{12} e^{i \phi}| = 
s_{12}c_{13} =|U_{e2}|$,  
$s^e_{12} \hat{s}_{23} = s_{13} = |U_{e3}|$, 
$c^e_{12} \hat{s}_{23} = s_{23}c_{13} = |U_{\mu 3}|$,
$\hat{c}_{23} =  c_{23}c_{13} = |U_{\tau 3}|$, etc. 
It is more difficult technically to demonstrate that 
for the elements $U_{\mu 1}$, $U_{\mu 2}$, 
$U_{\tau 1}$, $U_{\tau 2}$, but it can be 
easily checked numerically using, e.g.,  the best fit 
values of the angles  $\sin^2\theta_{12}$, 
$\sin^2\theta_{23}$ and $\sin^2\theta_{13}$ 
to determine numerically 
$\theta^e_{12}$, $\hat\theta_{23}$ and $\phi$, and 
correspondingly, $\delta$,
for each given value of $\theta^\nu_{12}$, 
and then using these ``data'' 
to calculate the absolute values of the indicated 
elements of the PMNS matrices given in 
eqs. (\ref{eq:VQ}) - (\ref{eq:Vpara}) 
and in eq. (\ref{Uthenuhat}).   
As a consequence, the PMNS matrix 
in eq. (\ref{Uthenuhat}) can be written as 
\begin{equation} 
\begin{array}{c}
\label{|U|phases}
U_{\rm PMNS} = \left(\begin{array}{ccc} 
|U_{e1}| e^{\ci\beta_{e1}} & |U_{e2}| e^{\ci\beta_{e2}} & 
|U_{e3}| e^{\ci \phi}\\[0.2cm]  
|U_{\mu 1}| e^{\ci\beta_{\mu 1}}& |U_{\mu 2}| e^{\ci\beta_{\mu 2}} & 
|U_{\mu 3}| e^{\ci \phi}
\\[0.2cm] 
|U_{\tau 1}| &  |U_{\tau2}|e^{- \ci \pi} 
 & |U_{\tau 3}| 
\\ 
\end{array}    
\right)\, Q_1\,Q_0\,,
\end{array} 
\end{equation}
%
where 
\begin{align}
\label{be1}
\beta_{e1} & = 
{\rm arg}\left (c^e_{12} c^\nu_{12} - 
 s^e_{12}\hat{c}_{23} s^\nu_{12} e^{\ci \phi} \right )\,,
\\[0.25cm]
\label{be2}
\beta_{e2} & = 
{\rm arg} \left ( c^e_{12} s^\nu_{12} 
+ s^e_{12}\hat{c}_{23} c^\nu_{12} e^{\ci \phi} \right )\,,
\\[0.25cm]
\label{bmu1}
\beta_{\mu 1} & = 
{\rm arg} \left ( - s^e_{12} c^\nu_{12}
 -  c^e_{12}\hat{c}_{23} s^\nu_{12} e^{\ci \phi} \right )\,,
\\[0.25cm]
\label{bmu2}
\beta_{\mu 2} & = 
{\rm arg} \left (- s^e_{12} s^\nu_{12} 
+  c^e_{12}\hat{c}_{23} c^\nu_{12} e^{i \phi} \right )\,.
\end{align}
%
The phases $\beta_{e1}$, $\beta_{e2}$, $\beta_{\mu1}$ and 
$\beta_{\mu 2}$ can be calculated for any of the specific 
values of $\theta^\nu_{12}$ of interest since, 
for a given $\theta^\nu_{12}$, the angles  
$\theta^e_{12}$, $\hat\theta_{23}$ and the phase $\phi$ 
can be determined from the values of 
the neutrino mixing parameters $\sin^2\theta_{12}$, 
$\sin^2\theta_{23}$ and $\sin^2\theta_{13}$. 
One has to remember that although $\cos\phi$ 
is uniquely determined, the sign of $\sin\phi$ 
cannot be determined using the current data.
Thus, two values of $\phi$, and correspondingly 
of the phases $\beta_{e1}$, $\beta_{e2}$, $\beta_{\mu1}$,
$\beta_{\mu 2}$ and $\delta$, are compatible with the data
 and have to be considered.

  As we know from the analysis in Section 3, the phase $\phi$ 
does not coincide with the Dirac phase $\delta$. 
It is not difficult to convince oneself that we have:
\begin{equation}
\delta = -\, \phi + \beta_{e1} + \beta_{e2}\,.
\label{dphibeta}
\end{equation}
%
Using eqs. (\ref{s2th13}) - (\ref{s2th12}), (\ref{be1}) 
and (\ref{be2}), it is rather straightforward 
to demonstrate, for instance, that 
$\sin (-\phi + \beta_{e1} + \beta_{e2}) = 
-\, \sin\phi \sin2\theta^\nu_{12}/\sin2\theta_{12} = \sin\delta$, 
where the last equality follows from eq. (\ref{sindsinphi}).   
The result given in eq. (\ref{dphibeta}) indicates what rearrangement 
of the phases in the PMNS matrix in eq. (\ref{|U|phases}) 
we have to perform in order to bring it to the standard 
parametrisation form:
\begin{equation} 
\begin{array}{c}
\label{Uphases}
U_{\rm PMNS} = P_2\, 
\left(\begin{array}{ccc} 
|U_{e1}| & |U_{e2}|  & |U_{e3}| e^{-\ci(-\phi + \beta_{e1} + \beta_{e2})}\\[0.2cm]  
|U_{\mu 1}| e^{\ci(\beta_{\mu 1} + \beta_{e2} -\phi)}& 
|U_{\mu 2}| e^{\ci(\beta_{\mu 2} + \beta_{e1} -\phi)} & |U_{\mu 3}|
\\[0.2cm] 
|U_{\tau 1}|e^{\ci\beta_{e2}} &  |U_{\tau2}|e^{\ci (\beta_{e1} - \pi)} 
 & |U_{\tau 3}| 
\\ 
\end{array}    
\right)\, Q_{2}\, Q_1\, Q_0\,,
\end{array} 
\end{equation}
%
where, as we have shown, 
$(-\phi + \beta_{e1} + \beta_{e2}) = \delta$ and 
\begin{align}
\label{P2}
P_2 &={\rm diag}(e^{\ci(\beta_{e1} + \beta_{e2})},e^{\ci \phi},1)\,,\\[0.25cm]
\label{Q2}
Q_2 &= {\rm diag} \left(e^{-\ci\beta_{e2}},e^{-\ci\beta_{e1}},1 \right)
= e^{-\ci\beta_{e2}}\,{\rm diag} \left(1,e^{\ci(\beta_{e2} - 
\beta_{e1})},e^{\ci\beta_{e2}} \right)\,.
\end{align}
%
The phases in the diagonal matrix $P_2$ are unphysical - they can be 
absorbed by the electron and muon fields in the weak charged lepton current.
The phases  $(\beta_{e2} - \beta_{e1})$ and $\beta_{e2}$ 
in the diagonal matrix 
$Q_2$ give contribution to the Majorana phases $\alpha_{21}/2$ and 
$\alpha_{31}/2$, respectively, while the common phase $(-\beta_{e2})$ in $Q_2$ 
is also unphysical and we will not keep it in our further 
analysis. One can show further (analytically or numerically) 
that we have:
\begin{align}
\label{mu1ph}
\beta_{\mu 1} + \beta_{e2} -\phi & = 
{\rm arg}(U_{\mu 1}) = 
{\rm arg}\left (-s_{12} c_{23} - c_{12} s_{23} s_{13} e^{\ci \delta} \right )\,,
\\[0.25cm]
\label{mu2ph}
\beta_{\mu 2} + \beta_{e1} -\phi & = 
{\rm arg}(U_{\mu 2}) =  
{\rm arg}\left ( c_{12} c_{23} - s_{12} s_{23} s_{13} e^{\ci \delta} \right )\,,
\\[0.25cm]
\label{tau1ph}
\beta_{e2} & = 
{\rm arg}(U_{\tau 1}) = 
{\rm arg} \left (s_{12} s_{23} - c_{12} c_{23} s_{13} e^{\ci \delta}\right )\,,
\\[0.25cm]
\label{tau2ph}
\beta_{e 1} & = 
{\rm arg}(U_{\tau 2}) + \pi =   
{\rm arg}\left[ \left (-c_{12} s_{23} 
- s_{12} c_{23} s_{13} e^{\ci \delta} \right )e^{\ci\pi} \right ]\,.
\end{align}
%
This implies that the matrix in eq. (\ref{Uphases}) 
is in the standard parametrisation form. 
Correspondingly, the Majorana phases 
$\alpha_{21}/2$ and $\alpha_{31}/2$ in the matrix $Q$  
in eq. (\ref{eq:VQ}) are determined 
by the phases in the matrix $\overline{Q} = Q_{2}Q_{1}Q_{0}$:
$Q = \overline{Q}$ and
\begin{equation}
\frac{\alpha_{21}}{2} = \beta_{e2} - \beta_{e1} +\frac{\xi_{21}}{2}\,,~~~  
\frac{\alpha_{31}}{2} = \beta_{e2} + \beta + \frac{\xi_{31}}{2}\,. 
\label{Majph2131}
\end{equation}
%

 The expressions we have obtained for the phases 
$\beta_{e1}$ and  $\beta_{e2}$, eqs. (\ref{be1}), (\ref{tau2ph})
and (\ref{be2}), (\ref{tau1ph}), are exact.
It follows from these expressions that 
the phases $\beta_{e1}$, $\beta_{e2}$ can be 
determined knowing the values of $\theta^e_{12}$ and $\phi$, or, 
alternatively, of $\delta$ and of 
$\theta_{12}$, $\theta_{23}$ and $\theta_{13}$. In what concerns 
the  phases  $\xi_{21}$ and $\xi_{31}$ in eq. (\ref{Majph2131}),
they are assumed to be fixed 
by the symmetry which determines the 
TBM, BM, GRA, etc. form of the matrix $\tilde{U}_{\nu}$.

  More specifically, the phases  $\beta_{e1}$, $\beta_{e2}$ can be 
calculated either using eqs. (\ref{be1}) and (\ref{be2}), 
or from eqs. (\ref{tau1ph}) and (\ref{tau2ph}).
It follows from eqs.  (\ref{tau1ph}) and (\ref{tau2ph}), in particular, 
that we have approximately
$|\sin\beta_{e1}| \cong 
\tan\theta_{12}\cot\theta_{23}\sin\theta_{13}|\sin\delta|\cong 
0.12|\sin\delta|$, and 
$|\sin\beta_{e2}| \cong 
\cot\theta_{12}\cot\theta_{23}\sin\theta_{13}|\sin\delta|\cong 
0.27|\sin\delta|$, where we have used the b.f.v. of 
$\sin^2\theta_{12}$, $\sin^2\theta_{23}$ and $\sin^2\theta_{13}$ 
quoted in eqs. (\ref{th12values}) - (\ref{th13values}).
These estimates imply that $\cos\beta_{e1}$ and 
 $\cos\beta_{e2}$ will have values close to 1.
Indeed, we get, e.g., utilising the values of $\cos\delta$ 
given in eqs. (\ref{cosdTBM}) - (\ref{cosdHG}) 
and the corresponding  b.f.v. of 
$\sin^2\theta_{12}$, $\sin^2\theta_{23}$ and 
$\sin^2\theta_{13}$:
\begin{align}
\label{cosbe1TBM}
{\rm TBM: } \qquad& \cos\beta_{e1} \cong  0.9929\,,~~~
 \beta_{e1} \cong \pm 6.81^\circ\,,
&\\[0.25cm]
\label{cosbe1BM}
{\rm BM~(LC): }  \qquad& \cos\beta_{e1} \cong  0.999\,,~~~
 \beta_{e1} \cong \pm 1.77^\circ\,,
&\\[0.25cm]
\label{cosbe1GRA}
{\rm GRA: } \qquad& 
\cos\beta_{e1} \cong  0.994\,,~~~
 \beta_{e1} \cong \pm 6.31^\circ\,,
& \\[0.25cm]
\label{cosbe1GRB}
{\rm GRB: } \qquad& 
\cos\beta_{e1} \cong  0.9927\,,~~~
\beta_{e1} \cong \pm 6.92^\circ\,,
& \\[0.25cm]
\label{cosbe1HG}
{\rm HG: }  \qquad& 
\cos\beta_{e1} \cong  0.995\,,~~~
 \beta_{e1} \cong \pm 5.79^\circ\,,
\end{align}
%
\begin{align}
\label{cosbe2TBM}
{\rm TBM: } \qquad& \cos\beta_{e2} \cong  0.976\,,~~~
 \beta_{e2} \cong \pm 12.64^\circ\,,
&\\[0.25cm]
\label{cosbe2BM}
{\rm BM~(LC): }  \qquad& \cos\beta_{e2} \cong  0.9989\,,~~~
 \beta_{e2} \cong \pm 2.58^\circ\,,
&\\[0.25cm]
\label{cosbe2GRA}
{\rm GRA: } \qquad& 
\cos\beta_{e2} \cong  0.973\,,~~~
 \beta_{e2} \cong \pm 13.24^\circ\,,
& \\[0.25cm]
\label{cosbe2GRB}
{\rm GRB: } \qquad& 
\cos\beta_{e2} \cong  0.977\,,~~~
\beta_{e2} \cong \pm 12.31^\circ\,,
& \\[0.25cm]
\label{cosbe2HG}
{\rm HG: }  \qquad& 
\cos\beta_{e2} \cong  0.975\,,~~~
 \beta_{e2} \cong \pm 12.91^\circ\,.
\end{align}
%
We see that the phases $\beta_{e1}$ and $\beta_{e2}$, 
with the exception of the BM (LC) case, 
take values approximately in the intervals 
$\pm (6^\circ - 7^\circ)$ and 
$\pm (12^\circ - 13^\circ)$, respectively.
For the phase difference $(\beta_{e2} - \beta_{e1})$, 
which contributes to the Majorana phase 
$\alpha_{21}/2$, we get taking into account that 
$\sin\beta_{e1}\sin\beta_{e2} < 0$:
\begin{align}
\label{cosbe21TBM}
{\rm TBM: } \qquad& \cos(\beta_{e2} - \beta_{e1}) \cong  0.943\,,~~~
 \beta_{e2}-\beta_{e1} \cong \pm 19.45^\circ\,,
&\\[0.25cm]
\label{cosbe21BM}
{\rm BM~(LC): }  \qquad& \cos(\beta_{e2} - \beta_{e1}) \cong  0.997\,,~~~
\beta_{e2} - \beta_{e1} \cong \pm 4.35^\circ\,,
&\\[0.25cm]
\label{cosbe21GRA}
{\rm GRA: } \qquad& \cos(\beta_{e2} - \beta_{e1}) \cong  0.942\,,~~~
 \beta_{e2} - \beta_{e1} \cong \pm 19.55^\circ\,,
& \\[0.25cm]
\label{cosbe21GRB}
{\rm GRB: } \qquad& \cos(\beta_{e2} - \beta_{e1}) \cong  0.944\,,~~~
\beta_{e2} - \beta_{e1}  \cong \pm 19.23^\circ\,,
& \\[0.25cm]
\label{cosbe21HG}
{\rm HG: }  \qquad& \cos(\beta_{e2} - \beta_{e1})\cong  0.947\,,~~~
 \beta_{e2}  - \beta_{e1}  \cong \pm 18.70^\circ\,.
\end{align}
%

 It follows from the results we have obtained that 
the contributions of the phases 
$2(\beta_{e2} - \beta_{e1})$ and $2\beta_{e2}$ 
to the Majorana phases $\alpha_{21}$ and $\alpha_{31}$ 
are practically negligible in the BM (LC) case.
In all other cases of the form of the matrix 
$\tilde{U}_{\nu}$ considered by us, 
TBM, GRA, GRB and HG, these contributions have to 
be taken into account. If the sign of $\sin\delta$ 
will be determined experimentally, the ambiguity 
in the signs of $\sin\beta_{e1}$,  $\sin\beta_{e2}$ 
and $\sin(\beta_{e2} - \beta_{e1})$ 
will be removed and $\beta_{e1}$, $\beta_{e2}$
and $(\beta_{e2} - \beta_{e1})$ will be uniquely 
determined. 

 We note that by writing,  
$2\beta_{e2} = \pm r_2^\circ$ 
and $2(\beta_{e2} -\beta_{e1})= \pm r_{21}^\circ$
we imply, in the convention used by us for the intervals in which 
the phases $\alpha_{21}$ and  $\alpha_{31}$ vary,
$2\beta_{e2} = ^{~+~}_{(-)}2r_2^\circ + 360^\circ k_2(')$ 
and 
$2(\beta_{e2} -\beta_{e1})= ^{~+~}_{(-)}2r_{21}^\circ + 360^\circ k_{21}(')$, 
$k_2,k_{21}=0,1$ ($k_2',k'_{21} = 1,2$), where $k_2 = 1$ ($k'_2 = 2$) 
and  $k_{21} = 1$ ($k'_{21} = 2$) 
has to be taken into account in certain  
cases \cite{EMSPEJP09} 
when the flavour neutrino Majorana mass term 
is generated by the type I seesaw mechanism \cite{seesaw}.

We will consider next the possibility to 
calculate also the phase $\beta = \gamma - \phi$ determined in 
eqs. (\ref{alphabeta}) and (\ref{gammaphi}).
We note first that the phase $\beta$ enters only in 
the expression for the Majorana phase $\alpha_{31}$.
The latter plays a subdominant role in a number of cases 
of processes, characteristic of the Majorana nature 
of massive neutrinos $\nu_j$. 
More specifically, the term involving the Majorana phase 
$\alpha_{31}$ gives a subdominant contribution 
in the $\betabeta$-decay rate in the cases of 
neutrino mass spectrum i) with inverted ordering (IO), 
corresponding to $\Delta m^2_{31(32)} < 0$, 
and ii) of quasi-degenerate (QD) type  
(see, e.g., \cite{PDG2014,WRodej10}), 
the reason being that the term of interest 
involves the suppression factor 
$\sin^2\theta_{13}\cong (0.023-0.024)$.
For the same reason the rate of
the process of radiative emission of two different 
Majorana neutrinos in atomic physics 
depends weakly on the Majorana phase $\alpha_{31}$ 
\cite{Dinh:2012qb}. The value of the phase $\alpha_{31}$ 
  plays important role, 
for example, for the prediction of the $\betabeta$-decay rate 
if neutrino mass spectrum is 
with normal ordering (NO) but is not quasi-degenerate,
i.e., if  $\Delta m^2_{31(32)} > 0$, $m_1 < (\ll) m_{2,3}$ 
and $m_1 \ltap \sqrt{\Delta m^2_{31}} \cong 0.05$ eV
(see, e.g., \cite{PDG2014}). 

  In the case of negligibly small $\theta^e_{23}$, as we have seen, 
$\gamma = - \psi + \pi$,  $\phi = - \psi$, and
$\beta = \pi$. In the ``counter-intuitive'' case \cite{FPR04} of 
$|\sin\theta^e_{23}| = 1$ we have $\gamma = \phi = - \omega$, and
$\beta = 0$. In these cases we get, e.g., for 
$(\xi_{21},\xi_{31}) = (0,0)$ using eqs. (\ref{cosbe2TBM}) - (\ref{cosbe21HG}):
\begin{align}
\nonumber
{\rm TBM: } \qquad& \alpha_{21}  \cong  \pm 38.90^\circ\,,~~~
 \alpha_{31} \cong \pm 25.28^\circ + 180^\circ~(0^\circ)\,,
&\\[0.25cm]
\nonumber
{\rm BM~(LC): }  \qquad& \alpha_{21}  \cong \pm 8.70^\circ\,,~~~
 \alpha_{31} \cong \pm 5.16^\circ + 180^\circ~(0^\circ) \,, 
&\\[0.25cm]
\nonumber
{\rm GRA: }  \qquad& \alpha_{21}  \cong \pm 39.10^\circ\,,~~~
 \alpha_{31} \cong \pm 26.48^\circ + 180^\circ~(0^\circ) \,,
& \\[0.25cm]
\nonumber
{\rm GRB: }  \qquad& \alpha_{21}  \cong \pm 38.46^\circ\,,~~~
 \alpha_{31} \cong \pm 24.62^\circ + 180^\circ~(0^\circ) \,,
& \\[0.25cm]
\nonumber
{\rm HG: }  \qquad& \alpha_{21}  \cong \pm 37.40^\circ\,,~~~
 \alpha_{31} \cong \pm 25.82^\circ + 180^\circ~(0^\circ) \,,
\end{align}
%
where the values (values in brackets) correspond to 
$\beta = \pi$ ($\beta = 0$).

  In the general case of non-negligible 
$\theta^e_{23}$ we get  from eq. (\ref{gammaphi}), 
using eq. (\ref{th23hat}):
\ba
\label{cosgamma}
\cos\gamma = \frac{-\,\cos\theta^e_{23}\cos\psi + \sin\theta^e_{23}\cos\omega}
{\sqrt{2} \sin\hat\theta_{23}}\,, \;\;\; 
\label{singamma}
\sin\gamma = \frac{\cos\theta^e_{23}\sin\psi - \sin\theta^e_{23}\sin\omega}
{\sqrt{2} \sin\hat\theta_{23}}\,,\\[0.30cm]
\label{cosphi}
\cos\phi = \frac{\cos\theta^e_{23}\cos\psi + \sin\theta^e_{23}\cos\omega}
{\sqrt{2} \cos\hat\theta_{23}}\,, \;\;\;
\label{sinphi}
\sin\phi = \frac{-\,\cos\theta^e_{23}\sin\psi - \sin\theta^e_{23}\sin\omega}
{\sqrt{2} \cos\hat\theta_{23}}\,. 
\ea
%
As it is not difficult to show using 
eqs. (\ref{cosgamma}) - (\ref{sinphi}),  
the phase $\beta$ depends on the 
phases $\psi$ and $\omega$ only via 
their difference $(\psi - \omega)$.
Indeed, we have:
\ba
\label{cosbeta}
\cos\beta 
= -\, \frac{\cos2\theta^e_{23}}{\sin2\hat\theta_{23}}\,,~~~
\sin\beta = \frac{\sin 2\theta^e_{23}}{\sin2\hat\theta_{23}}\,
\sin(\psi - \omega)\,,
\ea
where 
\begin{equation} 
\sin2\hat\theta_{23} = \left ( 1 - 
\sin^2 2\theta^e_{23}\,\cos^2(\psi -\omega) \right )^{\frac{1}{2}}\,.
\label{2the23psiomega}
\end{equation}
%
Thus, we have two undetermined 
parameters $\theta^e_{23}$ and  $(\psi - \omega)$, 
which are constrained by 
their relation to, e.g.,  $\sin^2\hat\theta_{23}$, 
whose value is known:
\begin{equation} 
2\,\sin\theta^e_{23}\,\cos\theta^e_{23}\,\cos(\psi -\omega) = 
1 - 2\,\sin^2\hat\theta_{23}\,.
\label{the23psiomega}
\end{equation}
%
This constraint reduces the number of the 
unknown parameters in terms of which 
the phase $\beta$ is expressed to one. 
The sign of $\sin(\psi -\omega)$ is also 
undetermined. Obviously, it is impossible 
to determine the phase $\beta$
without some additional input. 
In what follows we will exploit several possibilities.

 The first possibility corresponds to 
the phase $\psi$ or the phase $\omega$ 
having one of the following specific values:
$0,\pi/2,\pi$ and $3\pi/2$. 
In any of these cases the phase $\gamma$ 
is determined (up to a possible sign ambiguity 
either of $\sin\gamma$ or of $\cos\gamma$) 
by the phase $\phi$, which allows to 
determine also the phase $\beta$ 
(again up to a possible sign ambiguity 
of $\cos\beta$ or of $\sin\beta$).
This possibility 
is realised in certain models of neutrino 
mixing based on discrete flavour symmetries.

To be more specific, assume first 
that $\psi = 0$. In this case we get from 
eqs. (\ref{cosgamma}) - (\ref{sinphi}):
\ba 
\label{singppsi0} 
\sin\gamma = \sin\phi \, \frac{\cos\hat\theta_{23}}
{\sin\hat\theta_{23}}\,,~~~|\sin\gamma| \leq 1\,,\hspace{1.6cm} \\[0.30cm]
\cos\phi \cos\hat\theta_{23} + \cos\gamma \sin\hat\theta_{23} = 
\sqrt{2}\sin\theta^e_{23}\cos\omega\,.
\label{cosgppsi0}
\ea 
%
It is clear from eq. (\ref{singppsi0}) 
that the value of  $\sin\gamma$ can be determined 
knowing the values of $\sin\phi$ and of $\cot\hat\theta_{23}$,
independently of the values of $\theta^e_{23}$ and $\omega$.
This, obviously, allows to find also $|\cos\gamma|$, 
but not the sign of $\cos\gamma$.
If, however, the inequality 
$\sqrt{2}|\sin\theta^e_{23}\cos\omega| < |\cos\phi \cos\hat\theta_{23}|$ 
is fulfilled, eq. (\ref{cosgppsi0}) would allow to correlate the sign of 
 $\cos \gamma$ with the sign of $\cos\phi$ 
and thus to determine $\gamma$ for a given $\phi$:
we would have  $\cos\gamma < 0$ if $\cos\phi >0$, 
and $\cos\gamma > 0$ for $\cos\phi < 0$. 
In the case of 
$\sqrt{2}|\sin\theta^e_{23}\cos\omega| > |\cos\phi \cos\hat\theta_{23}|$, 
the sign of $\cos\gamma$ will coincide with the sign of 
$\sin\theta^e_{23}\cos\omega$, and if the latter cannot be fixed, 
the two possible signs of $\cos\gamma$ have to be considered.

  In a similar way we find that if $\psi = \pi/2$ we have:
\ba 
\label{singppsipi2} 
\cos\gamma = \cos\phi \, \frac{\cos\hat\theta_{23}}
{\sin\hat\theta_{23}}\,,\hspace{4.35cm} \\[0.30cm]
\sin\phi \cos\hat\theta_{23} + \sin\gamma \sin\hat\theta_{23} = 
-\, \sqrt{2}\sin\theta^e_{23}\sin\omega\,.
\label{cosgppsipi2}
\ea 
%
These relations hold in the model with 
$T^\prime$ family symmetry proposed in 
\cite{Girardi:2013sza}~ 
\footnote{We correct two typos eqs. (3.71) and (3.72) 
in \cite{Girardi:2013sza}: 
i) the factor $\sin\hat\theta_{23}/\cos\hat\theta_{23}$ in the 
r.h.s. of eq. (3.71) should be replaced by the inverse one, 
$\cos\hat\theta_{23}/\sin\hat\theta_{23}$, 
and ii) the factor $1/\sqrt{2}$ in the r.h.s. of eq. (3.72) 
should be replaced by $\sqrt{2}$, i.e., 
eqs. (3.71) and (3.72) in \cite{Girardi:2013sza} 
should coincide respectively with eqs. (\ref{singppsipi2}) and 
(\ref{cosgppsipi2}) given above.}.
Now the value of  $\cos\gamma$ can be determined 
knowing the values of $\cos\phi$ and of $\cot\hat\theta_{23}$,
independently of the values of $\theta^e_{23}$ and $\omega$.
This allows to find also $|\sin\gamma|$, 
leaving the sign of $\sin\gamma$ undetermined.
Depending on the relative magnitude of 
the terms $|\sin\phi \cos\hat\theta_{23}|$
and $|\sqrt{2}\sin\theta^e_{23}\sin\omega|$, 
the sign of $\sin\gamma$ will be anti-correlated 
either with the sign of $\sin\phi$, or 
with the sign of $\sin\theta^e_{23}\sin\omega$.
In the latter case both signs of $\sin\gamma$ 
have to be considered if the  sign of 
$\sin\theta^e_{23}\sin\omega$ is undetermined. 

 Similar results can be obtained if 
$\psi = \pi$ or $3\pi/2$, 
or if $\omega$ has one of the four values 
$0$, $\pi/2$, $\pi$ and $3\pi/2$.

  One finds $\beta = \pi + 2\pi k$, $k=0,1$, if 
the equality $\psi = \omega$ holds. This possibility is realised 
in a scheme considered in \cite{Duarah:2012bm}, in which 
also  the phases $\xi_{21}$  and $\xi_{31}$ are fixed:\\
$\xi_{21} = 0$ and  $\xi_{31}/2 = - \psi = - {\rm arg}[ 
(c^e_{23} - s^e_{23})e^{-\ci \phi}/(c^e_{23} + s^e_{23})]$, 
where  $\theta^e_{23}$ is determined from
eq. (\ref{the23psiomega}) in which one has to set
$\cos(\psi -\omega) = 1$.

 Further, it follows from eq. (\ref{cosbeta}) that if 
$|\sin\theta^e_{23}|$ (or $|\cos\theta^e_{23}|$)  
is known, that will allow to determine 
$\cos\beta$ and, correspondingly, 
$|\sin\beta|$. If, for instance, 
$|\sin\theta^e_{23}| = 0.2$, for the ``best fit'' value of 
$\sin^2\hat\theta_{23} = 0.438$ we find:
$\cos\beta \cong -\,0.919$, and
thus $\beta = 156.8^\circ$ or $203.2^\circ$.

 In a general analysis in which 
one attempts to reproduce the values of the 
three neutrino mixing parameters 
$\sin^2\theta_{12}$, $\sin^2\theta_{23}$ and  $\sin^2\theta_{13}$
in the cases of the TBM, BM, GRA, etc. forms of the matrix 
$\tilde{U}_{\nu}$ with the help of the ``correcting'' matrix 
$(\tilde{U}_e)^\dagger \Psi = 
R_{12}(\theta^e_{12})R_{23}(\theta^e_{23})\Psi$,  
the four parameters 
$\theta^e_{12}$, $\theta^e_{23}$, $\psi$ and $\omega$ 
will have to satisfy three constraints.
This implies that the values of any two parameters, say, 
$\theta^e_{23}$ and $(\psi - \omega)$, 
will have to be correlated 
\footnote{The author would like to thank W. Rodejohann and 
He Zhang for useful discussions of this point.
}.  
In addition, $\theta^e_{23}$ and $(\psi - \omega)$
have to satisfy the constraint given in 
eq. (\ref{th23hat}). This can allow to limit 
significantly the range of 
possible values of, or even to determine, 
$|\sin\theta^e_{23}|$. As a consequence, 
$\cos\beta$ (and therefore $|\sin\beta|$)
will either be constrained 
to lie in a relatively narrow interval, 
or its value will be determined, 
which will lead to a similar information 
about the phase $\beta$ 
(up to the possible ambiguity related 
to the two possible signs of $\sin\beta$).
Such an analysis, however, is outside the scope of the 
present investigation; we intend to perform it elsewhere.

\vspace{-0.2cm}
%
\section{Implications for $\betabeta$- Decay}
%

We will discuss next briefly the implications 
of the results we have obtained on the Dirac and Majorana phases 
for the predictions of the effective Majorana mass in 
$\betabeta$-decay (see, e.g., \cite{BiPet87,WRodej10}):
\begin{equation}
\meff=\left| m_1\, (c_{12}c_{13})^2 
+ m_2\, (s_{12}c_{13})^2\, e^{i\alpha_{21}}
 + m_3\, s^2_{13}\,e^{i(\alpha_{31} - 2\delta)} \right|\,,
\label{meff}
\end{equation}
%
where $m_j \geq 0$, $j=1,2,3$, are the masses of the three light 
Majorana neutrinos. As is well known, 
the existing data do not allow one to 
determine the sign of 
$\Delta m^2_{31(32)}$ and
the two possible signs of
$\Delta m^2_{31(2)}$ correspond to two 
types of neutrino mass spectrum.
In the widely used convention of numbering 
the neutrinos $\nu_j$ with definite mass 
in the two cases (see, e.g., \cite{PDG2014})
we shall also employ, the two spectra read:\\
{\it i) spectrum with normal ordering (NO)}:
$0 \leq m_1 < m_2 < m_3$, $\Delta m^2_{31(32)} >0$,
$\Delta m^2_{21} > 0$,
$m_{2(3)} = (m_1^2 + \Delta m^2_{21(31)})^{\frac{1}{2}}$; \\~~
{\it ii) spectrum with inverted ordering (IO)}:
$0 \leq m_3 < m_1 < m_2$, $\Delta m^2_{32(31)}< 0$, 
$\Delta m^2_{21} > 0$,
$m_{2} = (m_3^2 + \Delta m^2_{23})^{1\over{2}}$, 
$m_{1} = (m_3^2 + \Delta m^2_{23} - \Delta m^2_{21})^{\frac{1}{2}}$.\\
The values of $\Delta m^2_{21} > 0$ and $\Delta m^2_{31} > 0$
($\Delta m^2_{32} < 0$) in the NO (IO) case were determined with 
relatively high precision in the global analyses 
of the neutrino oscillation data and read 
\cite{Capozzi:2013csa}:
\begin{equation}
\label{delt21values}
 (\Delta m^2_{21})_{\rm BF} = 7.54 \times 10^{-5}~{\rm eV^2}\,, 
 \Delta m^2_{21} = (6.99 - 8.18) \times 10^{-5}~{\rm eV^2}\,;
\end{equation}
\begin{equation}
\label{del312values}
(|\Delta m^2_{31(32)}|)_{\rm BF} = 2.48~(2.44) \times 10^{-3}~{\rm eV^2}\,,  
~~~|\Delta m^2_{31(32)}| = (2.26~(2.21) - 2.70~(2.65)) \times 10^{-3}
~{\rm eV^2}\,,
\end{equation}
%
where we have given the best fit values and the $3\sigma$ 
allowed ranges of $\Delta m^2_{21}$ and $|\Delta m^2_{31}|$
($|\Delta m^2_{32}|$). Thus, we have, in particular, 
$\Delta m^2_{21}/|\Delta m^2_{31(32)}| \cong 0.03$.

 Consider  the case of IO neutrino mass spectrum.
Expressing $m_{1.2}$ in terms of $m_3$, 
$\Delta m^2_{21}$ and $\Delta m^2_{23} >0$ in eq. (\ref{meff}), 
and taking into account 
the fact that $\Delta m^2_{21} \ll \Delta m^2_{23}$, 
we get:
\begin{equation}
\meff \cong \sqrt{m^2_3 + \Delta m^2_{23}}
\left|c^2_{13} \left (c^2_{12} + s^2_{12} e^{i\alpha_{21}}\right )
- \frac{1}{2}\,
\frac{\Delta m^2_{21} (c_{12}c_{13})^2}
{m^2_3 + \Delta m^2_{23}}
 + s^2_{13}\,\frac {m_3 e^{i(\alpha_{31} - 2\delta)}}
{\sqrt{m^2_3 + \Delta m^2_{23}}}\right|\,.
\label{meffIO1}
\end{equation}
%
It follows from eq. (\ref{th12values}) that at $3\sigma$ we have:
$|c^2_{12} + s^2_{12}\, e^{i\alpha_{21}}| \geq 0.28$.
Taking into account the result on $\sin^2\theta_{13}$ 
quoted in eq. (\ref{th13values}), 
it is clear that the term 
$\propto s^2_{13}m_3$ in eq. (\ref{meffIO1}) 
is at least by a factor of 10 smaller in absolute value than 
$|c^2_{12} + s^2_{12}\, e^{i\alpha_{21}}|$.
The term $\propto \Delta m^2_{21}$ in eq. (\ref{meffIO1})
does not exceed approximately 0.01.
Thus, up to corrections which are not larger than 10\%, 
$\meff$ in the case of IO spectrum is given by \cite{BPP1}:
\begin{equation}
\meff \cong \sqrt{m^2_3 + \Delta m^2_{23}}\,
\left|c^2_{12} + s^2_{12}\, e^{i\alpha_{21}}\right| = 
 \sqrt{m^2_3 + \Delta m^2_{23}}\,
\left (1 -  
\sin^22\theta_{12}\,\sin^2\frac{\alpha_{21}}{2}\right)^{\frac{1}{2}}\,. 
\label{meffIO2}
\end{equation}
%

The expression for $\meff$ in the case of 
QD neutrino mass spectrum (see, e.g., \cite{PDG2014}),
$m_1\cong m_2\cong m_3$, 
$m^2_{1,2,3} >> |\Delta m^2_{31(32)}|$, implying   
$m_0 \equiv {\rm min}(m_j) \gtap 0.1$ eV, 
has a similar form up to corrections 
$\sim |\Delta m^2_{31(32)}|/m^2_0$ \cite{BPP1}: 
\begin{equation}
\meff \cong m_0\,\left|c^2_{12} + s^2_{12}\, e^{i\alpha_{21}}\right| = 
m_0 \, \left (1 -  
\sin^22\theta_{12}\,\sin^2\frac{\alpha_{21}}{2}\right)^{\frac{1}{2}}\,. 
\label{meffQD}
\end{equation}
%

It follows from eqs. (\ref{meffIO2}), (\ref{Majph2131}) and 
(\ref{cosbe21TBM}) - (\ref{cosbe21HG}) that 
for $\xi_{21} = 0$ and the best fit values of the 
neutrino mixing angles, $\meff$ will deviate little from  
the maximal possible value corresponding to the IO spectrum, 
$\meff \cong \sqrt{m^2_3 + \Delta m^2_{23}}$,  
since for all cases considered 
$\sin^2(\alpha_{21}/2) = \sin^2(\beta_{e2} - \beta_{e1}) \ltap 0.11$.
If, however, $\xi_{21} = \pi$, then 
$\sin^2(\alpha_{21}/2) = \cos^2(\beta_{e2} - \beta_{e1})$ 
and $\meff$ can be expected to 
be closer to its minimal possible value of
$\meff \cong \sqrt{m^2_3 + \Delta m^2_{23}}~\cos2\theta_{12}$.
Using $\sin^22\theta_{12} = 0.85$ 
(corresponding to $\sin^2\theta_{12} = 0.308$ 
for which $\cos2\theta_{12}= 0.39)$ and 
the values of  $\cos(\beta_{e2} - \beta_{e1})$     
given in eqs. (\ref{cosbe21TBM}) - (\ref{cosbe21HG}), 
we get: $\meff \cong C_a~\sqrt{m^2_3 + \Delta m^2_{23}}$,
$a=TBM,BM(LC),GRA,GRB,HG$, where
$C_{TBM} \cong 0.49$, $C_{BM(LC)} \cong 0.39$, 
$C_{GRA} \cong 0.49$, $C_{GRB} \cong 0.49$ and 
$C_{HG} \cong 0.48$. Thus, in the BM (LC) case 
$\meff$ is minimal, while in the other cases 
$\meff$ is approximately half 
of its maximal value.
For any other value of $\xi_{21}$, 
the prediction for $\meff$ 
for a given symmetry case will lie between 
those quoted for  $\xi_{21} = 0$ and 
and  $\xi_{21} = \pi$.
For the TBM, GRA, GRB and HG symmetry 
mixing, this implies that 
$0.49\sqrt{m^2_3 + \Delta m^2_{23}} \ltap 
\meff \leq \sqrt{m^2_3 + \Delta m^2_{23}}$,
while for the BM (LC) mixing case, 
$0.39 \sqrt{m^2_3 + \Delta m^2_{23}} \ltap 
\meff \leq \sqrt{m^2_3 + \Delta m^2_{23}}$, 
where the numerical factors correspond to the 
best fit values of $\sin^2\theta_{12}$, 
 $\sin^2\theta_{23}$ and  $\sin^2\theta_{13}$.
Similar results are valid for the 
QD neutrino mass spectrum.

 One can use the same method to obtain predictions 
for $\meff$ in the case of non-QD neutrino mass spectrum 
with normal ordering in the cases when the phase $\beta$ 
is  known.

\vspace{-0.3cm}
\section{Summary and Conclusions}

 We have applied the approach developed in 
ref. \cite{Marzocca:2013cr} to 
obtaining predictions for the 
Dirac and  Majorana CP violation phases 
in the neutrino mixing (PMNS) matrix. 
The approach is based on the 
fact that the PMNS matrix 
$U_{\text{PMNS}} = U_e^{\dagger}\, U_{\nu} = 
(\tilde{U}_{e})^\dagger\, \Psi \tilde{U}_{\nu} Q_0$, 
where $U_{e}$ ($\tilde{U}_{e}$) 
and $U_{\nu}$ ($\tilde{U}_{\nu} Q_0$) 
result respectively from the diagonalisation 
of the charged lepton and neutrino mass matrices, 
$\tilde{U}_{e}$ and $\tilde{U}_{\nu}$ are $3\times3$ 
unitary (CKM-like) matrices, and
$\Psi$ and  $Q_0$ are diagonal phase matrices  
containing, in general, two physical 
CP violation phases each.
The phases in $Q_0$, $\xi_{21}/2$ and $\xi_{31}/2$, contribute to 
the Majorana phases $\alpha_{21}/2$ and $\alpha_{31}/2$, 
present in the standard parametrisation of the PMNS matrix 
(see eq. (\ref{eq:VQ})). The CPV phases in
$\Psi$ can originate from the charged lepton sector 
($U_e^{\dagger} = (\tilde{U}_{e})^\dagger\, \Psi$),
or from the neutrino sector  
($U_{\nu} = \Psi \tilde{U}_{\nu} Q_0$), 
or  can receive contributions from both sectors.
We have considered a number of 
different forms of  $\tilde{U}_{\nu} = 
\tilde{U}_{\nu}(\theta^{\nu}_{12},\theta^{\nu}_{23},\theta^{\nu}_{13},\delta^\nu)$
associated with a variety of flavour symmetries,
for which $\theta^{\nu}_{13} = 0$ 
(and thus one can set $\delta^\nu = 0$) and 
$\theta^{\nu}_{23} = -\pi/4$:
i) bimaximal (BM) ($\theta^{\nu}_{12} = \pi/4$),
and ii) tri-bimaximal (TBM) 
($\theta^{\nu}_{12} = \sin^{-1} (1/\sqrt{3})$) forms, 
the forms corresponding iii) to the conservation of the 
lepton charge $L' = L_e - L_\mu - L_{\tau}$ (LC) 
($\theta^{\nu}_{12} = \pi/4$), 
iv) to golden ratio type A (GRA) mixing
with $\sin^2\theta^{\nu}_{12} = (2 + r)^{-1} \cong 0.276$, 
$r$ being the golden ratio,  $r = (1 +\sqrt{5})/2$, 
v) golden ratio type B (GRB) mixing, 
with   $\sin^2\theta^{\nu}_{12} = (3 - r)/4 \cong 0.345$,
and vi) to hexagonal (HG) mixing,
in which $\theta^{\nu}_{12} = \pi/6$. 
The TBM, BM and GRA
special forms of $\tilde{U}_{\nu}$, for instance, 
can be obtained from specific discrete family 
symmetries in the lepton sector (see, e.g., 
\cite{Everett:2008et,GRAM,GRBM,HGM,King:2013eh,Alta:2010ab,Tani:2010cd}). 
In the cases of symmetry forms of $\tilde{U}_{\nu}$
considered, the phases in the matrix  
$\Psi =
{\rm diag}(1,\text{e}^{-\ci \psi},\text{e}^{-\ci \omega})$,
generate the Dirac phase $\delta$ in the 
(standard parametrisation of the) PMNS matrix
and, as we have shown, give rise to contributions 
to the Majorana phases  $\alpha_{21}/2$ and $\alpha_{31}/2$.
The minimal form of $\tilde{U}_e$, 
in terms of angles it contains, that can provide the requisite 
corrections to $\tilde{U}_{\nu}$ so that 
reactor, atmospheric and solar neutrino mixing angles 
$\theta_{13}$, $\theta_{23}$ and  $\theta_{12}$ 
have values compatible with the current data,
including a possible sizable deviation of $\theta_{23}$ 
from $\pi/4$, is a product of two orthogonal matrices 
describing rotations in the 12 and 23 planes, 
$R_{12}(\theta^e_{12})$ and $R_{23}(\theta^e_{23})$. 
Two orderings of the 
12 and the 23 rotations in $\tilde{U}_e$ are possible:
``standard'' with 
$\tilde{U}_e = R_{23}(\theta^e_{23}) R_{12}(\theta^e_{12})$, 
and ``inverse'' with  
$\tilde{U}_e = R_{12}(\theta^e_{12})R_{23}(\theta^e_{23})$.  
The ``standard'' ordering is related to the 
hierarchy of the charged lepton masses, 
$m_e^2 \ll m^2_{\mu} \ll m^2_{\tau}$,  
and is a common feature of the overwhelming majority 
of the existing models of the charged lepton 
masses and the associated mixing.
In the present article we have analysed 
only the more interesting  case of  ``standard''
ordering. In this case the Dirac CP violation phase 
$\delta$, present in the PMNS matrix $U$, 
is shown to satisfy a new sum rule, 
eq. (\ref{cosdthnu}), by which 
$\cos\delta$ is expressed in terms of the 
angle $\theta^{\nu}_{12}$ of $\tilde{U}_{\nu}$ and
the three angles $\theta_{12}$, 
$\theta_{23}$ and $\theta_{13}$ of the PMNS matrix.
Within the approach employed the sum rule we have 
derived is exact and is 
a generalisation of the 
sum rule found in \cite{Marzocca:2013cr} 
for the TBM and BM (LC) forms of 
$\tilde{U}_\nu$. This allowed us to 
obtain predictions for $\delta$ and the 
$J_{CP}$ factor, which controls the magnitude of the CP 
violation effects in neutrino oscillations,
in the cases of GRA, GRB and HG forms of 
$\tilde{U}_{\nu}$; predictions for $\delta$ and $J_{CP}$   
for the TBM and BM (LC) forms 
of $\tilde{U}_\nu$ were obtained
in  \cite{Marzocca:2013cr}.
Although the $\cos\delta$ is determined without 
sign ambiguity, the sign of  $\sin\delta$
cannot be fixed using the current data, 
which leads to a two-fold (sign) ambiguity 
in the value of $\delta$.
The indicated results on $\delta$ and  the 
$J_{CP}$ factor are given in eqs. (\ref{cosdTBM}) - (\ref{cosdHG}) and 
eqs. (\ref{JCPTBM}) - (\ref{JCPHG}). 
They have been derived for the best fit values 
of the neutrino mixing parameters $\sin^2\theta_{12}$, 
$\sin^2\theta_{23}$ and $\sin^2\theta_{13}$. 
It follows from these results that:\\ 
i) $\delta \cong 1.59\pi$ or $0.41\pi$ in the GRA case;
ii) $\delta \cong 1.45\pi$ or $0.55\pi$ in the GRB case; and
iii) $\delta \cong 1.64\pi$ or $0.36\pi$ in the HG case.\\
In the TBM and BM (LC) cases we have \cite{Marzocca:2013cr}:
iv) TBM:  $\delta \cong 1.47\pi$ or $0.53\pi$;
v) BM (LC): $\delta \cong 1.07\pi$ or $0.93\pi$.
Thus, in the TBM, GRA, GRB and HG cases, relatively large CP 
violation effects in neutrino oscillations 
are predicted ($|J_{CP}| \cong (0.031 - 0.034)$), 
while in the BM (LC) case the indicated CP violation 
effects are suppressed. Distinguishing between the 
TBM, BM (LC), GRA, GRB and HG forms of $\tilde{U}_{\nu}$  
requires a measurement of $\cos\delta$ or a 
relatively high precision measurement of $J_{\rm CP}$. 

  We have considered also the case of 
$|\sin\theta^e_{23}| \ll 1$ (Section 4), 
analysing first the possibility of negligibly small 
$|\sin\theta^e_{23}|$ (subsection 4.1). 
For $\theta^e_{23} = 0$ we have
$\tilde{U}_e = R_{12}(\theta^e_{12})$. This case has 
been analysed by many authors in the past
(see, e.g., refs. \cite{GTani02}-\cite{Duarah:2012bm}). 
If  $\theta^e_{23} = 0$, as is well known,  
$\sin^2\theta_{23}$ can deviate only by 
$0.5\sin^2\theta_{13}$ from 0.5. 
The phase $\omega$ is unphysical. 
Now the exact sum rule of interest involves 
the cosine of the Dirac phase $\delta$ 
and the angles  $\theta^{\nu}_{12}$, 
$\theta_{12}$ and $\theta_{13}$ (eq. (\ref{cosd23e0})). 
A similar sum rule can be obtained for the cosine of the
phase  $\phi = - \psi$, eq. (\ref{s2th12se230phi}), 
which is related to, but does not coincide with, 
$\delta$ (the exact relation between 
$\cos\delta$ and $\cos\phi$ for 
arbitrary $\theta^{\nu}_{12}$ is given in
\footnote{The exact relation between $\delta$ ($\cos\delta$) and 
$\phi$ ($\cos\phi$) in the case of  $\theta^e_{23} \neq 0$, 
in which  $\phi \neq - \psi$ and $\phi$ is defined in 
eq. (\ref{gammaphi}), is given in eq. (\ref{dphibeta}) 
(eq. (\ref{cosdcosphi})).} 
eq. (\ref{cosdcosphi23e0})).
We have derived exact, leading order and next-to-leading order 
sum rules for both $\cos\phi$ and $\cos\delta$, 
eqs. (\ref{s2th12se230phi}), (\ref{s2th12se230appr2}),  
(\ref{s2th12se230appr1}) and 
eqs. (\ref{s2th12se230delta}), (\ref{s2th12se230deltaappr2}), 
(\ref{s2th12se230deltaappr1}), respectively.
The leading order sum rules (\ref{s2th12se230appr2}) and 
(\ref{s2th12se230deltaappr2}) 
are shown to be equivalent to the sum rules 
for $\cos\delta$ and $\cos\phi$ given in eqs. 
 (\ref{AntKingphi}) and (\ref{AntKingdelta}).
For arbitrary $\theta^{\nu}_{12}$, 
the leading order sum rule 
(\ref{AntKingdelta}) was proposed 
in \cite{Antusch:2005kw}. 
In  ref. \cite{Antusch:2007rk} it was suggested 
that the sum rule (\ref{AntKingdelta}) 
should be used to obtain the value of $\cos\delta$ 
using the experimentally determined values of 
$\sin^2\theta_{12}$ and $\sin\theta_{13}$, e.g., 
in the case of the TBM form of 
$\tilde{U}_{\nu}$. 
We have shown that the  
sum rule (\ref{AntKingdelta}) 
is the leading order approximation of the 
exact sum rule (\ref{cosdthnu}), derived in 
Section 3. We have also shown that 
in the cases of TBM,  GRA, GRB and HG forms of 
$\tilde{U}_{\nu}$, and for the current 
best fit value of $\sin^2\theta_{12}$,
the leading order sum rule 
(\ref{AntKingdelta}) is not consistent 
with the approximation employed to derive it. 
A consistent application of the corrections 
in the indicated cases
leads to $\cos\delta = \cos\phi = 0$.
As a consequence, the next-to-leading order 
corrections to (\ref{AntKingdelta}), or to the 
equivalent sum rule (\ref{s2th12se230deltaappr2}), 
derived in eq. (\ref{s2th12se230deltaappr1}) 
(and in eq. (\ref{s2th12se230appr1}) for 
$\cos\phi$), are significant and should be taken 
into account. For the TBM GRA, GRB and HG forms of 
$\tilde{U}_{\nu}$, the predictions for 
$\cos\delta$ (and $\cos\phi$) 
derived using the exact sum rule eq. (\ref{cosd23e0}) (eq. 
(\ref{s2th12se230phi})), or the 
next to leading order sum rule 
eq. (\ref{s2th12se230deltaappr1}) 
(eq. (\ref{s2th12se230appr1})), 
differ by factors of (1.2 - 1.6) 
from the predictions obtained from 
the leading order sum 
rule eq. (\ref{AntKingdelta}) 
(eq.  (\ref{AntKingphi})), 
or the equivalent one
eq. (\ref{s2th12se230deltaappr2}) 
(eq. (\ref{s2th12se230appr2})).
As we have shown in subsection 4.2, 
this difference can be further amplified 
by an additional factor of 1.2 
by the next-to-leading order correction due 
to $\theta^e_{23}\neq 0$, 
$\sin\theta^e_{23}\ll 1$, if 
$\sin^2\theta_{23} \cong 0.4$.
Using the exact sum rules 
eqs. (\ref{cosdthnu}) and 
(\ref{cphi}) leads for $\theta^e_{23}\neq 0$ 
to practically the same results
respectively for $\cos\delta$ and $\cos\phi$
as the next-to-leading order sum rules 
eq. (\ref{s2th12deltaappr3}) and eq. (\ref{s2th12phiappr3}).
We have shown also that 
the leading order sum rule 
(\ref{AntKingdelta}) provides  
a rather accurate prediction 
for $\cos\delta$ only in the case of 
BM (LC) form of the matrix  
$\tilde{U}_{\nu}$.

  In Section 5 we have analysed the possibility 
to obtain predictions for the values of the Majorana 
phases  $\alpha_{21}/2$ and $\alpha_{31}/2$ in the 
PMNS matrix. We have shown that   
$\alpha_{21}/2 = \beta_{e2} - \beta_{e1} + \xi_{21}/2$ and  
$\alpha_{31}/2 = \beta_{e2} + \beta + \xi_{31}/2$, 
where $\xi_{21}$ and $\xi_{31}$ are the phases 
of the matrix $Q_0$, and $\beta_{e1}$, $\beta_{e2}$ and 
$\beta$ are real calculable phases.
In many theories and models of neutrino mixing
the values of the phases  $\xi_{21}$ and $\xi_{31}$ 
are fixed by the form of the neutrino 
Majorana mass term, which is dictated by the 
chosen discrete (or continuous) flavour symmetry
or on phenomenological grounds. 
Typical values of $\xi_{21}/2$ and $\xi_{31}/2$ are 
0, $\pi/2$ and $\pi$~ 
\footnote{In the model with $T^\prime$ flavour symmetry 
in the lepton sector 
constructed in \cite{Girardi:2013sza}, for instance, 
$\xi_{21}$ and $\xi_{31}$ can take two sets of values:
$(\xi_{21},\xi_{31}) = (0,0)$ and $(0,\pi)$.}. 
Within the approach adopted in the present article,
the phases $\beta_{e1}$ and $\beta_{e2}$
can be calculated exactly for each 
of the five symmetry forms of $\tilde{U}_\nu$ 
considered by us. We have first derived exact 
analytic expressions for  $\beta_{e1}$ and $\beta_{e2}$ 
in terms of the three neutrino mixing angles, 
$\theta_{12}$, $\theta_{23}$, $\theta_{13}$, 
and the Dirac phases $\delta$ (eqs. (\ref{tau2ph}) and (\ref{tau1ph})).
Given $\theta_{12}$, $\theta_{23}$, $\theta_{13}$ 
and $\theta^{\nu}_{12}$ (i.e., the symmetry form of $\tilde{U}_\nu$),  
these expressions allow to get predictions for the values 
of  $\beta_{e1}$ and $\beta_{e2}$. We give 
such predictions for  $\beta_{e1}$, $\beta_{e2}$
and  $(\beta_{e2} - \beta_{e1})$
for each of the five symmetry 
forms of  $\tilde{U}_\nu$ considered 
using the the best fit values of 
$\sin^2\theta_{12}$, $\sin^2\theta_{23}$ 
and $\sin^2\theta_{13}$ 
(eqs. (\ref{cosbe1TBM}) - (\ref{cosbe21HG})).
In what concerns the phase $\beta$ entering 
into the expression for the Majorana phase 
$\alpha_{31}/2$, we have discussed a number of cases in which it 
can be calculated exactly. 

  Finally, in Section 6 we have analysed the implications 
of the results obtained on the leptonic CPV phases for the 
predictions of the effective Majorana mass 
in $\betabeta$-decay. This was done on the examples of the 
neutrino mass spectra with inverted ordering and 
of quasi-degenerate type.

  The predictions for the leptonic CP violation phases in the 
PMNS neutrino mixing matrix derived in the present 
article will be tested in the 
experiments on CP violation in neutrino oscillations 
and possibly in the neutrinoless double beta decay 
experiments.

\vspace{-0.3cm}
\section*{Note Added.} After this study was completed, 
results of an updated global analysis of the neutrino oscillation data
were published in \cite{Capozzi:2013csav2}, in which the latest 
T2K data on $\sin^2\theta_{23}$ \cite{T2K:1403.1532},  
$\sin^2\theta_{23} = 0.514+0.055/-0.056~(0.511 \pm 0.055)$ 
for the NO (IO) neutrino mass spectrum, 
were taken into account. As a consequence, 
the authors of \cite{Capozzi:2013csav2} 
find a somewhat larger central value of 
$\sin^2\theta_{23}$ than the one used by us  
in the numerical predictions for the Dirac and Majorana phases, 
namely  $\sin^2\theta_{23} = 0.437~(0.455)$ in the NO (IO) case. 
At the same time, the MINOS collaboration 
finds for the best fit value of  
$\sin^2\theta_{23} = 0.41$, performing 
a 3-neutrino oscillation analysis of 
their data \cite{MINOS:3nu1403}.
Obviously, high precision measurement of  $\sin^2\theta_{23}$ 
is lacking at present. Our numerical predictions for the values of the 
Dirac and Majorana phases should be updated when a sufficiently 
precise determination of  $\sin^2\theta_{23}$ will be available.
However, if  $\sin^2\theta_{23}$ is found to lie in the interval 
(0.40 - 0.50), the numerical predictions obtained in 
this study will not change significantly.

\vspace{0.2cm}
{\bf{\large Acknowledgements.}}
The author would like to thank 
I. Girardi, A. Titov, W. Rodejohann and 
He Zhang for useful discussions.
This work was supported in part also by the European Union FP7 
ITN INVISIBLES (Marie Curie Actions, PITN-GA-2011-289442-INVISIBLES),
by the INFN program on Theoretical Astroparticle Physics (TASP),
by the research grant  2012CPPYP7 ({\it  Theoretical Astroparticle Physics})
under the program  PRIN 2012 funded by the Italian Ministry of Education, University and Research (MIUR), and by the World Premier International Research Center
Initiative (WPI Initiative), MEXT, Japan.


\vspace{-0.3cm}
\begin{thebibliography}{99}

\bibitem{Marzocca:2013cr}
  D.~Marzocca, S.~T.~Petcov, A.~Romanino and M.~C.~Sevilla,
  JHEP {\bf 1305} (2013) 073.


\bibitem{PDG2014}
K. Nakamura and S.~T. Petcov,
in K.~A.~Olive {\it et al.}  
(Particle Data Group),
Chin. Phys. C {\bf 38} (2014) 090001.

\bibitem{LBLFuture13}
S.K. Agarwalla {\it et al.}, arXiv:1312.6520;
C. Adams {\it et al.}, arXiv:1307.5700;
A. de Gouvea {\it et al.}, arXiv:1310.4340.


\bibitem{BHP80} S.M.~Bilenky, J.~Hosek and S.T.~Petcov,
              Phys.\ Lett. B {\bf 94} (1980) 495.

\bibitem{EMSPEJP09}
  E.~Molinaro and S.~T.~Petcov,
  Eur.\ Phys.\ J.\  C {\bf 61} (2009) 93.

\bibitem{Ibarra:2011xn}
  A.~Ibarra, E.~Molinaro and S.~T.~Petcov,
Phys. Rev. D {\bf 84} (2011) 013005.   


\bibitem{LW81} L. Wolfenstein,  Phys. Lett. B {\bf 107} (1981) 77;
S.M.\ Bilenky, N.P.\ Nedelcheva and
S.T.\ Petcov,  Nucl. Phys. B\textbf{ 247} (1984) 61;
B. Kayser, Phys. Rev. D {\bf 30} (1984) 1023.

\bibitem{Lang87} P. Langacker {\it et al.},  
 Nucl. Phys. B {\bf 282} (1987) 589.

\bibitem{BiPet87} S.~M.~Bilenky and S.~T.~Petcov,
   Rev. Mod. Phys.  {\bf 59} (1987) 671.

\bibitem{BPP1} S.M. Bilenky, S. Pascoli and S.T. Petcov,
      Phys.\ Rev. D{\bf 64} (2001) 053010; 
S.T. Petcov, 
 Physica Scripta {\bf T121} (2005) 94. 
%
\bibitem{WRodej10} W. Rodejohann, 
 Int. J. Mod. Phys. E {\bf 20} (2011) 1833.

\bibitem{Yoshimura:2006nd}
  M.~Yoshimura,
  Phys.\ Rev.\ D {\bf 75} (2007) 113007.


\bibitem{Dinh:2012qb}
  D.~N.~Dinh {\it et al.},
  Phys.\ Lett.\ B {\bf 719} (2013) 154.
 
\bibitem{PPY03} S. Pascoli {\it et al.}, 
Phys. Lett. B {\bf 564} (2003) 241;
  S.~T.~Petcov, T.~Shindou and Y.~Takanishi,
  Nucl. Phys. B {\bf 738} (2006) 219.

\bibitem{Pascoli:2006ie}
  S.~Pascoli, S.~T.~Petcov and A.~Riotto,
  Phys. Rev. D {\bf 75} (2007) 083511, and
%
 Nucl. Phys. B {\bf 774} (2007) 1.

\bibitem{Capozzi:2013csa}
  F.~Capozzi {\it et al.},
 Phys.\  Rev. \ D {\bf 89} (2014) 093018.
 
\bibitem{GonzalezGarcia:2012sz}
 M. C. Gonzalez-Garcia {\it et al.}, 
 JHEP {\bf 12} (2012) 123;
the updated results obtained after the TAUP2013 International 
 Conference (held in September of 2013) are posted at the URL 
 www.nu-fit.org/?q=node/45.  


\bibitem{T2K1113th13} K. Abe {\it et al.}, 
Phys. Rev. Lett. {\bf 112} (2014) 061802.

\bibitem{DBay1013th13}  F.P. An {\it et al.},
Phys. Rev. Lett. {\bf 112} (2014) 061801.

\bibitem{Girardi:2013sza}
  I.~Girardi {\it et al.},
  JHEP {\bf 1402} (2014) 050.

\bibitem{Luhn:2013lkn}
  C.~Luhn,
  Nucl.\ Phys.\ B {\bf 875} (2013) 80;
   S.~F.~King, T.~Neder and A.~J.~Stuart,
  Phys.\ Lett.\ B {\bf 726} (2013) 312;
  A.~D.~Hanlon, S.~-F.~Ge and W.~W.~Repko,
  Phys.\ Lett.\ B {\bf 729} (2014) 185;
  B.~Dasgupta and A.~Y.~Smirnov,
  arXiv:1404.0272;
  G.~-J.~Ding and Y.~-L.~Zhou,
  arXiv:1404.0592.

\bibitem{Shimizu:2014ria}
  Y.~Shimizu and M.~Tanimoto,
  arXiv:1405.1521.

\bibitem{TBM}
  P.~F.~Harrison, D.~H.~Perkins and W.~G.~Scott,
  Phys.\ Lett.\ B {\bf 530} (2002) 167;
  Phys.\ Lett.\ B {\bf 535} (2002) 163;
  Z.~Z.~Xing,
  Phys.\ Lett.\ B {\bf 533} (2002) 85;
  X.~G.~He and A.~Zee,
  Phys.\ Lett.\ B {\bf 560} (2003) 87;
see also
L.~Wolfenstein,
  Phys.\ Rev.\ D {\bf 18} (1978) 958.

\bibitem{SPPD82} S.T. Petcov, Phys.\ Lett.\ B {\bf 110} (1982) 245.

\bibitem{BM}
F.~Vissani, 
[arXiv:{hep-ph/9708483}];
V.~D.~Barger {\it et al.}, 
Phys.\ Lett.\ B {\bf 437} (1998) 107;
A.~J.~Baltz, A.~S.~Goldhaber and M.~Goldhaber,
Phys.\ Rev.\ Lett.\  {\bf 81} (1998) 5730.
%

\bibitem{GTani02}
 C.~Giunti and M.~Tanimoto,
Phys.\ Rev.\ D {\bf 66} (2002) 113006;
see also: 
C.~Giunti and M.~Tanimoto,
Phys. Rev. D {\bf 66} (2002) 053013.

\bibitem{FPR04}
P.H. Frampton, S.T. Petcov and W. Rodejohann,
Nucl. Phys. B {\bf 687} (2004) 31.

\bibitem{SPWR04}
S.T. Petcov and W. Rodejohann,
Phys. Rev. {\bf D71} (2005) 073002.

\bibitem{Romanino:2004ww}
  A.~Romanino,
  Phys. Rev.  D {\bf 70} (2004) 013003.

\bibitem{Hochmuth:2007wq}
  K.~A.~Hochmuth, S.~T.~Petcov and W.~Rodejohann,
  Phys.\ Lett.\ B {\bf 654} (2007) 177.

\bibitem{Marzocca:2011dh}
 D.~Marzocca {\it et al.}, 
JHEP {\bf 11}  (2011) 009.

\bibitem{Alta} G.~Altarelli, F.~Feruglio and I.~Masina,
  Nucl. Phys. B {\bf 689} (2004) 157;
 I.~Masina,
  Phys.\ Lett.\  B {\bf 633} (2006) 134.

\bibitem{King2005}
  S.~F.~King,
   JHEP {\bf 0508} (2005) 105;

\bibitem{Antusch:2005kw}
  S.~Antusch and S.~F.~King,
  Phys.\ Lett.\ B {\bf 631} (2005) 42;

\bibitem{Antusch:2007rk}
  S.~Antusch {\it et al.}, 
  JHEP {\bf 0704} (2007) 060;

\bibitem{Antusch:2011qg}
  S.~Antusch and V.~Maurer,
  Phys. Rev. D {\bf 84} (2011) 117301;
A. Meroni {\it et al.}, 
 Phys. Rev. D {\bf 86} (2012) 113003;
  S.~Antusch {\it et al.}, 
  Nucl. Phys. B {\bf 866} (2013) 255.

\bibitem{Chen:2009gf}
  M.-C.~Chen and K.~T.~Mahanthappa,
  Phys.\ Lett.\ B {\bf 681} (2009) 444;
  M.-C.~Chen {\it et al.}, 
  JHEP {\bf 1310} (2013) 112.


\bibitem{Duarah:2012bm}
  C.~Duarah, A.~Das and N.~N.~Singh,
 arXiv:1210.8265. 


\bibitem{Chao:2011sp}
  W.~Chao and Y.~-j.~Zheng,
  JHEP {\bf 1302} (2013) 044;
  G.~Altarelli {\it et al.}, 
  JHEP {\bf 1208} (2012) 021;
  G.~Altarelli, F.~Feruglio and L.~Merlo,
  [arXiv:{1205.5133}]; 
  F.~Bazzocchi and L.~Merlo,
  [arXiv:{1205.5135}]; 
  S.~Gollu, K.~N.~Deepthi and R.~Mohanta,
  [arXiv:{1303.3393}]. 



\bibitem{Everett:2008et}
  L.~L.~Everett and A.~J.~Stuart,
  Phys.\ Rev.\ D {\bf 79} (2009) 085005.
 

\bibitem{GRAM}
  Y.~Kajiyama, M.~Raidal and A.~Strumia,
  Phys.\ Rev.\ D {\bf 76} (2007) 117301.

\bibitem{GRBM} 
  W.~Rodejohann,
  Phys.\ Lett.\ B {\bf 671} (2009) 267;
  A.~Adulpravitchai, A.~Blum and W.~Rodejohann,
  New J.\ Phys.\  {\bf 11} (2009) 063026.


\bibitem{HGM}
  C.~H.~Albright, A.~Dueck and W.~Rodejohann,
  Eur.\ Phys.\ J.\ C {\bf 70} (2010) 1099.
  [arXiv:1004.2798 [hep-ph]].


\bibitem{King:2013eh}
  S.~F.~King and C.~Luhn,
  Rept.\ Prog.\ Phys.\  {\bf 76} (2013) 056201.


\bibitem{Alta:2010ab}
G. Altarelli and F. Feruglio, 
Rev.\ Mod.\ Phys. {\bf 82} (2010) 2701.


\bibitem{Tani:2010cd} 
 H. Ishimori {\it et al.},
Prog.\ Theor.\ Phys.\ Suppl. {\bf 183} (2010) 1.

\bibitem{CarlMCC} C. H. Albright and M.-C. Chen,
Phys. \ Rev. D {\bf 74} (2006) 113006.

\bibitem{Kile:2014kya}
  J.~Kile {\it et al.}, 
  Phys.\ Rev.\ D {\bf 90} (2014) 013004;
  C.~C.~Li and G.~J.~Ding,
  arXiv:1408.0785 [hep-ph];
  C.~C.~Li and G.~J.~Ding,
  arXiv:1408.0785 [hep-ph];
  S.~Antusch {\it et al.},
  arXiv:1405.6962 [hep-ph].

\bibitem{Hall:2013yha}
  L.~J.~Hall and G.~G.~Ross,
  JHEP {\bf 1311} (2013) 091;
  Z.~Liu and Y.~L.~Wu,
  Phys.\ Lett.\ B {\bf 733} (2014) 226;
  S.~K.~Garg and S.~Gupta,
  JHEP {\bf 1310} (2013) 128;
  S.~Gollu, K.~N.~Deepthi and R.~Mohanta,
  Mod.\ Phys.\ Lett.\ A {\bf 28} (2013) 31, 1350131.

\bibitem{PKSP3nu88}
P.I.~Krastev and S.~T.~Petcov,
  Phys.\ Lett.\  B {\bf 205} (1988) 84.

\bibitem{King:2014nza}
  S.~F.~King {\it et al.},
  New J.\ Phys.\  {\bf 16} (2014) 045018.

\bibitem{Girardi:2014xyz} I. Girardi, A. Titov and S.T. Petcov, 
arXiv:1410.8056.

\bibitem{seesaw}
P.~Minkowski,
  Phys.\ Lett.\ B {\bf 67} (1977) 421;
T. Yanagida in {\it Proc. of the Workshop on Unified Theory and
Baryon Number of the Universe}, KEK, Japan, 1979;
M. Gell-Mann, P. Ramond and R. Slansky, 
talk at the Sanibel conference, 
Feb. 1979, 
and Print 80-0576, published in {\it Supergravity} 
(North Holland, Amsterdam 1979);
S.L.Glashow, Cargese Lectures (1979);
R.~N.~Mohapatra and G.~Senjanovic,
Phys. Rev. Lett.  {\bf 44} (1980) 912.

\bibitem{Capozzi:2013csav2}
  F.~Capozzi {\it et al.},
  arXiv:1312.2878v2 (May 5, 2014).

\bibitem{T2K:1403.1532} K. Abe {\it et al.}, 
arXiv:1403.1432.

\bibitem{MINOS:3nu1403} P. Adamson {\it et al.}, 
Phys.  Rev. Lett. {\bf 112} (2014) 191801. 

\end{thebibliography}
\end{document}